\documentclass[useAMS,usenatbib]{mn2e}
\usepackage{amsmath,fleqn,graphicx,amssymb,xcolor}
\usepackage{multirow}
\renewcommand{\[}{\begin{equation}}
\renewcommand{\]}{\end{equation}}
\def\p{\partial}\def\i{{\rm i}}

\def\rd{}
\let\boldgrk=\gkvecten
\let\boldgrksc=\gkvecseven

\def\gkthing#1{{\mathchoice%
	{\hbox{{\boldgrk\char#1}}}
	{\hbox{{\boldgrk\char#1}}}
	{\hbox{{\boldgrksc\char#1}}}
	{\hbox{{\boldgrksc\char#1}}}}}

\def\vdelta{\gkthing{14}}

{\newif\ifnotend
\notendtrue
\def\veclist{ABCDEFGHIJKLMNOPQRSTUVWXYZabcdefghijklmnopqrstuvwxyz.}
\def\top#1#2.{#1}
\def\tail#1#2.{#2.}
\loop\expandafter\xdef\csname v\expandafter\top\veclist\endcsname%
{{\noexpand\bf\expandafter\top\veclist}}
\edef\veclist{\expandafter\tail\veclist}
\if\veclist.\notendfalse\fi\ifnotend\repeat}
\def\d{{\rm d}}

\def\Gyr{\,\mathrm{Gyr}}
\def\Myr{\,\mathrm{Myr}}
\def\kpc{\,\mathrm{kpc}}

\def\kms{\,\mathrm{km\,s}^{-1}}

\def\msun{\,{\rm M}_\odot}\def\lsun{\,{\rm L}_\odot}

\def\pc{\,\mathrm{pc}}
\def\e{\mathrm{e}}

\def\fracj#1#2{{\textstyle{#1\over#2}}}

\def\rms{\textsc{rms}}

\title[Disc distortion revisited]
{Disc distortion revisited}

\author[James Binney]{
  James Binney$^1$\thanks{E-mail: binney@thphys.ox.ac.uk}\\  
  $^1$Rudolf Peierls Centre for Theoretical Physics, Clarendon Laboratory,
  Parks Road, Oxford, OX1 3PU, UK
}

\begin{document}
\maketitle

\begin{abstract}
We revisit the dynamics of razor-thin, stone-cold, {\rd self-gravitating
discs}. By recasting
the equations into standard cylindrical coordinates, the {\rd linearised} vertical dynamics of
an exponential disc can be followed {\rd for several gigayears on a laptop} in a
few minutes. An initially warped disc rapidly evolves into a flat inner
region and an outward-propagating spiral corrugation wave that rapidly winds
up and would quickly thicken a disc with non-zero radial velocity dispersion.
The Sgr dwarf galaxy generates a similar warp in the Galactic disc as it
passes through pericentre, and the warp generated by the dwarf's last
pericentre {\rd$\sim35\Myr$} ago is remarkably similar to the warp traced by the
Galaxy's HI disc. The resemblance to the observed warp is fleeting but its
timing is perfect. For the adopted parameters the amplitude of the model warp
is a factor 3 too small, but there are several reasons for this being so. The
marked flaring of {\rd our Galaxy's low-$\alpha$ disc} just outside {\rd the
solar circle} can be explained as a legacy of earlier pericentres.

\end{abstract}

\begin{keywords}
  Galaxy:
  kinematics and dynamics -- galaxies: kinematics and dynamics -- methods:
  numerical
\end{keywords}

\section{Introduction} \label{sec:intro}

The discs of spiral galaxies are typically remarkably planar inside
three disc
scale-lengths $R_\d$ but frequently become warped further out. Early
observations at 21 cm revealed this structure in the case of the Milky Way
\citep{Burke1957,Kerr1957}, and 21-cm observations with the Westerbork Aperture
Synthesis Radio Telescope revealed similar structure in a significant
fraction of nearby spiral galaxies \citep{Bosma1978}. \cite{Briggs1990} summarised and
systematised the phenomenology of the galactic warps. 

Data regarding the planarity of stellar discs are much less abundant for
several reasons. One issue is that a galaxy's stellar disc is much thicker
than its HI disc, so small deviations from planarity can be hidden in the
disc's thickness. Moreover, the mean velocity of a stellar disc is much
harder to measure with precision than the mean velocity of the HI disc, so
the velocities associated with warps are harder to detect in the stellar
disc. These {\rd issues} are least acute in the case of the Milky Way, but
historically the interpretation of data for the Milky Way has been made
uncertain by uncertain distances to stars. With the advent of the third data
release by ESA's Gaia mission \citep{GaiaDR3general} our ability to probe the vertical
dynamics of the Milky Way has dramatically improved. This turn of events
calls for a corresponding improvement in the dynamical theory of warped
discs.

The standard path to deep understanding in physics is usually linear
theory. Given that the most massive parts of galactic discs are planar to a
very good approximation, we naturally seek a linear theory of discs' vertical
perturbations. The problem cannot be approached by direct analogy with spiral
structure, that is by linearising the stellar disc's distribution function
$f(\vx,\vv)$, because the disc is thin, so the unperturbed distribution
function (DF), $f_0$, changes
by order of itself when the vertical coordinate $z$ changes by the disc's
scale height, which is  comparable to the amplitude of even a small warp.

{\rd Since the problem is not accessible to  Eulerian perturbation theory,}
\citealt[(hereafter HT69)]{HunterToomre1969} developed a {\rd linearised
Lagrangian} theory of
vertical disturbances of a razor-thin disc by modelling the unperturbed disc
as a system of $N$ perfectly circular, co-planar orbits and allowing orbits
to develop small inclinations.  {\rd With $h(R)$ denoting the amplitude of the
vertical excursions of orbits at radius $R$, they derived $2N$ equations of
motion linear in the small parameter $h/R$.} They investigated the normal
modes of this system, hoping to find discrete normal modes that resembled
observed warps. They concluded, however, that the system's spectrum tends to
a continuum as $N\to\infty$ whenever the disc has a realistically smooth
decline in density at large radii.

When a system's spectrum forms a continuum, perturbations of the system will
always be spread among modes that cover a non-negligible range of
frequencies, with the consequence  that the perturbation will fade away as
the phases of individual modes drift towards a random distribution. Since the
pioneering work of HT69 we have recognised that all stellar
system have continuous spectra and that this fact underpins the  dissipative
nature of their dynamics that is familiar from N-body simulations
\citep[e.g.][]{LauBinney2021a}.

Given the continuous nature of the spectra established by HT69
straightforward integration of the equations of motion offers a more
profitable route to understanding disc dynamics than normal-mode analysis.
HT69 performed some integrations, but they examined only an infinitesimal
fraction of the interesting possibilities, for two main reasons: (i) their equations
related to a coordinate system that is not well suited to exponential discs,
and (ii) they did not include a dark halo because they were writing before
dark haloes became accepted features of galactic structure.

When dark haloes had become established, the normal-mode approach to disc
dynamics was revived in the expectation that a flattened dark halo would
permit the existence of a non-trivial and isolated `tilt' mode that might
explain observed warps \citep{Toomre1983,SparkeCasertano1988}. An isolated
tilt mode can be found if the dark halo's potential is assumed to be fixed,
but when the back-reaction of a perturbed dark halo is included in the disc's
dynamics, the tilt mode dissolves and warps phase-mix away as they do in an
isolated disc \citep{NelsonTremaine1995,Binneyetal1998}. 

The practical impossibility of long-lived warp configurations leads one to
investigate the evolution of warps in the time rather than the frequency
domain. Here we return to the approach pioneered by HT69 and too rarely
employed since. Instead of using the oblate-spheroidal coordinate system that
HT69 used, we show that the dynamical equations can be integrated cheaply in
standard cylindrical polar coordinates when advantage is taken of the Greens
function for Poisson's equation that was introduced by
\cite{CohlTohline1999}. The problem's equations are derived in
Section~\ref{sec:eqs} -- they differ slightly from those derived in HT69 as
is discussed in Appendices \ref{sec:HTdiscuss} and \ref{sec:appHTeqs}.  In
Section~\ref{sec:freeWarp} we investigate the evolution of a disc that starts
with a simple warp. In Section~\ref{sec:SgrWarp} we explore the effect that
the Sgr dwarf galaxy has had on the vertical structure of our Galaxy's disc.
Section~\ref{sec:discuss} discusses the successes, failures and limitations
of the HT69 approach to vertical disc dynamics.  Section~\ref{sec:conclude}
sums up and looks to the future. 

\section{The equations}\label{sec:eqs}

{\rd In this section we obtain the equations that govern the vertical dynamics of particles
that move on near-circular orbits under their mutual gravitational
attraction.}

By solving Laplace's equation by separation of variables in cylindrical
polar coordinates, it can be shown that \citep{CohlTohline1999}
\[\label{eq:CohlTohlineeqsa}
{1\over|\vx-\vx'|}={1\over\pi\sqrt{RR'}}\sum_{m=-\infty}^\infty 
Q_{|m|-1/2}(\chi)\e^{\i m(\phi-\phi')},
\]
where $Q$ is a Legendre function of the second kind and
\[\label{eq:CohlTohlineeqsb}
\chi(R,z,R',z')\equiv{R^2+R^{\prime2}+(z-z')^2\over2RR'}.
\]
We apply this expansion to the density distribution 
\[
\rho(R',\phi',z')=\Sigma(R')\delta(z'-h')
\]
 of a warped disc, where $\Sigma(R')$ is the unperturbed surface density and
$h'(R',\phi')$ is the vertical displacement
of the disc at $\vx'=(R',\phi')$. Executing the integral over $z'$ yields
\begin{align}\label{eq:PhifromQ}
&\Phi(\vx)=-G\int{\rd\d^2\vx'}\,\Sigma(\vx'){\delta(z'-h')\over|\vx-\vx'|}\cr
&\ =-{G\over\pi}\sum_m\int\d R'\sqrt{R'\over
R}\Sigma(R')\int\d\phi'\,\e^{\i m(\phi-\phi')}Q_{|m|-1/2}\bigg|_{z'=h'}.
\end{align}
The vertical acceleration generated by $\Phi$ at $\vx=(R,\phi,h)$ is
\begin{align}\label{eq:beforePull}
a_z(R,h)&=-{\p\Phi\over\p z}\bigg|_{z=h}
={G\over\pi}\sum_m\int\d R'\sqrt{R'\over
R}\Sigma(R')\cr
&\quad\times\int\d\phi'\,\e^{\i m(\phi-\phi')}{\d
Q_{|m|-1/2}\over\d\chi}{h-h'\over RR'}.
\end{align}
We are seeking an expression for $a_z$ that's linear in {\rd$(h-h')/\sqrt{RR'}$, so we can
neglect the dependence of $\d Q/\d\chi$ on the small quantity
$(h-h')/\sqrt{RR'}$
by evaluating $\chi$ with $z-z'=\epsilon_z$, a distance that's much smaller
than $\sqrt{RR'}$} that will be discussed
below. This done, $\d Q/\d\chi$ becomes
independent of $\phi'$ and can be taken out of the $\phi'$ integral along
with $\e^{\i m\phi}/RR'$. We are left with {\rd the integral} 
\[
\int\d\phi'\,\e^{-\i m\phi'}(h-h').
\]
The part involving $h$ vanishes unless $m=0$, when it yields
$2\pi h(R,\phi)$, while that
involving $h'$ yields $2\pi h_m(R')$, where
\[\label{eq:hFourier}
h_m(R)=\int_0^{2\pi}{\d\phi\over2\pi}\e^{-\i m\phi}h(R,\phi).
\]
Thus
\begin{align}\label{aVfirst}
a_z&(R,h)={2G\over R^{3/2}}\int{\d R'\over\surd{R'}}\Sigma(R')\cr
&\times\bigg({\d
Q_{-1/2}\over\d\chi}h(R,\phi)-\sum_{m}{\d
Q_{|m|-1/2}\over\d\chi}h_m(R')\e^{\i m\phi}\bigg).
\end{align}
Multiplying by $\e^{-\i m\phi}/2\pi$ and integrating over $\phi$
we obtain an expression for the Fourier coefficients of $a_z$
\begin{align}
a_{zm}&={2G\over R^{3/2}}\int{\d R'\over\surd R'}\Sigma(R')\cr
&\times\bigg({\d Q_{-1/2}\over\d\chi}h_m(R)-{\d
Q_{|m|-1/2}\over\d\chi}h_m(R')\bigg).
\end{align}
 As $R'\to R$, $\chi$ tends to $1+\frac12(\epsilon_z/R)^2$ and the
derivatives of $Q$ diverge like $(R/\epsilon_z)^2$. Nonetheless, $a_{zm}$ is
finite {\rd because the divergent parts of the terms proportional to $h_m(R)$
and $h_m(R')$ cancel}. We avoid singular integrands by subtracting $(\d
Q_{|m|-1/2}/\d\chi) h_m$ from the first term in the big bracket and adding it
to the second term. Then we can write
\begin{align}\label{eq:azeqa}
a_{zm}&=
N_m(R)h_m(R)\cr
&-\int{\d R'\over\surd R'}\,M_m(R,R')\big[h_m(R')-h_m(R)\big],
\end{align}
where
\begin{align}\label{eq:azeqb}
N_m(R)&\equiv{2G\over R^{3/2}}\int{\d R'\over\surd R'}\Sigma(R')
\bigg({\d Q_{-1/2}\over\d\chi}-{\d
Q_{|m|-1/2}\over\d\chi}\bigg)\cr
M_m(R,R')&\equiv{2G\over R^{3/2}}\Sigma(R')
{\d Q_{|m|-1/2}\over\d\chi}.
\end{align}
 Numerical evaluations show that both contributions to $a_{zm}$ become
independent of $\epsilon_z$ when the latter is sufficiently small: in the
integral for $N_m$, the divergencies of the derivatives of $Q_{-1/2}$ and
$Q_{|m|-1/2}$ cancel, while in the integral over $M_m$ the factor $h'-h$
suppresses the contribution from $R'\simeq R$ in which $\d Q/\d\chi$ depends
on $\epsilon_z$. 

We  obtain differential equations for $h_m(R,t)$ from the equation of
motion $\ddot z=a_z$ of a star's $z$ coordinate: $z$ is the value
of $h$ at the star's location, $(R,\phi=\phi_0+\Omega t)$, where
$\Omega(R)$ is the
circular frequency at the star's location. Thus
\[\label{eq:basichddoteq}
\ddot z=\Big({\p\over\p
t}+\Omega(R){\p\over\p\phi}\Big)^2h=a_z.
\]
The Fourier components of $h$ satisfy corresponding equations
\begin{align}\label{eq:warpeqs}
\ddot h_m+2\i m&\Omega\dot h_m-m^2\Omega^2h_m = N_mh_m\cr
&-\int{\d R'\over\surd R'}M_m(R,R')\big[h_m(R')-h_m(R)\big],
\end{align}
 where we have used equation (\ref{eq:azeqa}) for the acceleration.  Each
function $h_m(R)$ satisfies its own p.d.e., completely independent of the
other $h_m$.  Hence {\rd in this linear theory} we can investigate
independently the dynamics of corrugations that are axisymmetric ($m=0$), or
involve tilts ($m=1$) or have more elaborate angular structure.

From equation (\ref{eq:hFourier}) it follows that
\begin{align}\label{eq:zexplicit}
z&(t)=h(t,\phi_0+\Omega t)=h_0\cr
&+2\sum_{m>0}\Big(\Re(h_m)\cos[m\phi(t)]
-\Im(h_m)\sin[m\phi(t)]\Big).
\end{align}
In the important case $m=1$, {\rd the} line of nodes, in which rings cross the plane
$z=0$, lies along
\[\label{eq:lons}
\phi=\hbox{atan2}\big[\Im(h_1),\Re(h_1)\big].
\]

Differentiating equation (\ref{eq:zexplicit}) wrt time, we have
\begin{align}\label{eq:vz}
v_z&=2\sum_{m>0}\big[\Re(\dot h_m)\cos(m\phi)
-\Im(\dot h_m)\sin(m\phi)\big]\cr
&-2\sum_{m>0}m\Omega\big[\Re(h_m)\sin(m\phi)
+\Im(h_m)\cos(m\phi)\big].
\end{align}

The derivation above of the equations of motion (\ref{eq:warpeqs}) is
superficially straightforward and simpler than that given by Hunter \& Toomre
in their classic paper, but it skates over a very delicate point. The issue
is explained in Appendix \ref{sec:HTdiscuss}.

\subsection{Adding a bulge and dark halo}

The gravitational fields of real galaxies include important contributions
from spheroidal components such as bulges and dark halos, so we should add
the potentials of a bulge and dark halo to the disc's potential.  An
externally sourced gravitational field modifies equation \eqref{eq:warpeqs}
in two ways. First, it changes $\Omega(R)$, which should be the actual
circular frequency rather than just the frequency generated by the disc
itself.  Second, the external field adds a term $a_{3z}$ to the right of
equation \eqref{eq:basichddoteq} for the vertical acceleration contributed by
the external gravitational field. Given that the bulge and halo are
reflection-symmetric in the equatorial plane, the potential $\Phi_\e(R,z)$
that they generate must be an even function of $z$.  Hence its Taylor series
gives 
 \[\label{eq:haloPot}
\Phi_\e(R,z)=\Phi_\e(R,0)+\fracj12{\p^2\Phi_\e\over\p z^2}\bigg|_{z=0}z^2+\cdots
\]
 and to a good approximation we have that the corresponding vertical
acceleration is $a_{3z}=-Kz$, where {\rd$K\equiv\p^2\Phi_\e/\p z^2\big|_{z=0}$} is a positive
constant.

It is easy to show that in the case of a spherical {\rd external potential
$\Phi_\e$} and a corrugation with $m=1$, the additional acceleration $a_{3z}$
on the right of equation \eqref{eq:warpeqs} that one gets from equation
\eqref{eq:haloPot} exactly cancels the increase in $\Omega^2$ on the left of
equation \eqref{eq:warpeqs}\footnote{\rd In a spherical potential orbits do
not precess with the consequence that $\dot h=0$ must solve equation
\eqref{eq:warpeqs} when $\Sigma=0$.}, so the addition of a spherical potential
modifies equation \eqref{eq:warpeqs} only by increasing the magnitude of the
Coriolis term $2\i m\Omega\dot h_m$ that couples the real and imaginary parts
of $h_m$. The physical consequence of this change is to increase the rate at
which rings precess. 

\subsection{Tidal excitation}\label{sec:tidaleqs}

As many authors have remarked \citep[e.g.][]{McMillan2022}, the Sagittarius
dwarf galaxy is probably a major contributor to excitation of the Galactic
warp. We can use equations \eqref{eq:warpeqs} to explore this possibility by
adding a suitable tidal forcing to the equations' right sides before
integrating them from $h=\dot h=0$. We work in the freely-falling frame of the
Galactic centre. In this frame the effective gravitational field at $\vx$ due
to a perturber with a spherical potential $\Psi$ is
\begin{align}\label{eq:tidalg}
\vg(t,\vx)&=-\nabla\Psi_{(\vX-\vx)}+\nabla\Psi_\vX\cr
&= {\d\Psi\over\d r}\bigg|_{|\vX-\vx|}{\vX-\vx\over|\vX-\vx|}-
{\d\Psi\over\d r}\bigg|_{|\vX|}{\vX\over|\vX|},
\end{align}
 where $\vX(t)$ is the perturber's location relative to the Galactic centre.
On a circle of radius $R$ in the plane, the vertical component of $\vg$ can
be Fourier decomposed into the sum
\[\label{eq:tidalFT}
\ve_z\cdot\vg(R,\phi,t)=\sum_mg_m(R,t)\e^{\i m\phi}
\]
{\rd Vertical} excitation of the disc by the dwarf is then obtained by adding $g_m$ to the right
side of equation \eqref{eq:warpeqs}. {\rd Unfortunately, the radial and azimuthal
excitations have to be ignored.} 

\subsection{Integrating the equations of motion}\label{sec:integrating}

To integrate the equations of motion (\ref{eq:warpeqs}) we introduce a radial grid $R_n$ by the rule
\[
R_n=R_\d\times
\begin{cases}0.05\delta&\hbox{if }$n=0$\cr
\sinh(n\delta)&\hbox{otherwise},
\end{cases}
\]
so $R_{n+1}-R_n\simeq R_\d\delta$ near the centre but far out we have
$R_n\simeq \frac12R_\d\e^{n\delta}$ and
$R_{n+1}/R_n\simeq\e^\delta$. {\rd The grid employed had 300 nodes and extended to
$10R_\d$ with $\delta=\sinh^{-1}(10/299)$.}
We force the innermost ring to remain in the
plane by holding $h_m(R_0)=0$. 

By numerical integration we compute the values of $N_m(R)$ at each grid
point and store them. For each grid value of $R$ we similarly compute and
store the integral in equation (\ref{eq:warpeqs})
over each interval $R'\in(R_j,R_{j+1})$ with $h'-h$ replaced by the $n$th
power of
$t=(R'-R_j)/(R_{j+1}-R_j)$ for $n=0,\ldots,3$.
 At each timestep a cubic spline is fitted
to the values of $h_m$ on the grid and then the integral in equation
\eqref{eq:warpeqs} is evaluated as $\sum_{in}M_i^nH_{in}$, where $H_{in}$ is the
coefficient of $t^n$ for the interval starting at $R_i$ in the spline
representation of $h_m$. The seventh-order Runge-Kutta algorithm is used to
advance the $h_m$ in time. 

\begin{figure}
\centerline{\includegraphics[width=.8\hsize]{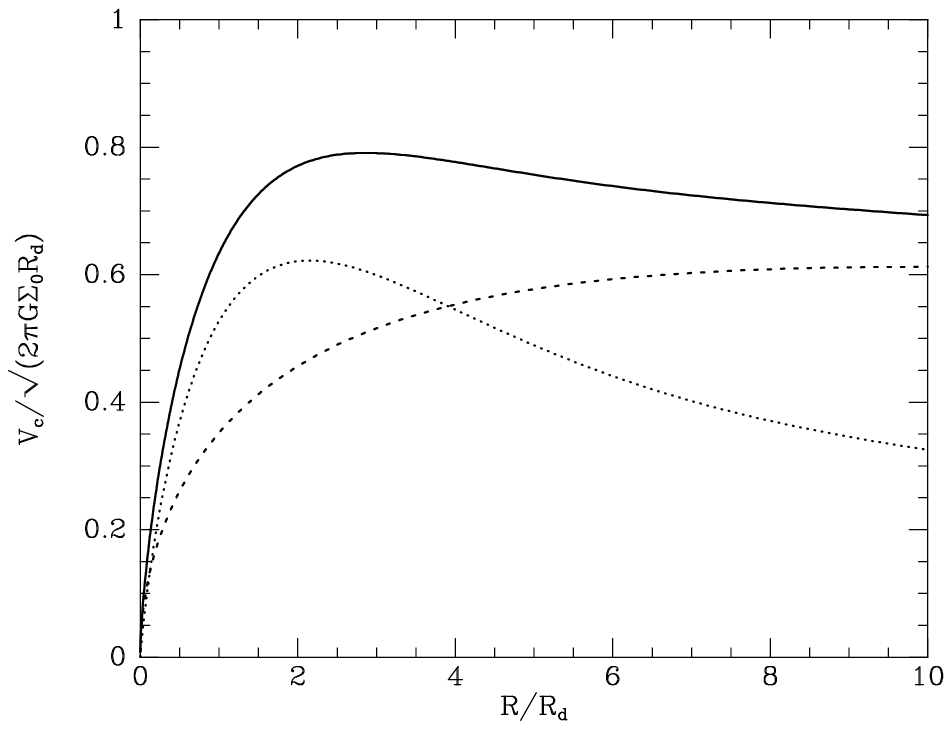}}
\caption{The full curve shows the circular-speed of the galaxy model used in Section
\ref{sec:freeWarp}. The
contributions of the dark halo and the disc are shown by the dashed and
dotted curves, respectively.}\label{fig:freeMWVc}
\end{figure}

\begin{figure*}
\centerline{
\includegraphics[width=.32\hsize]{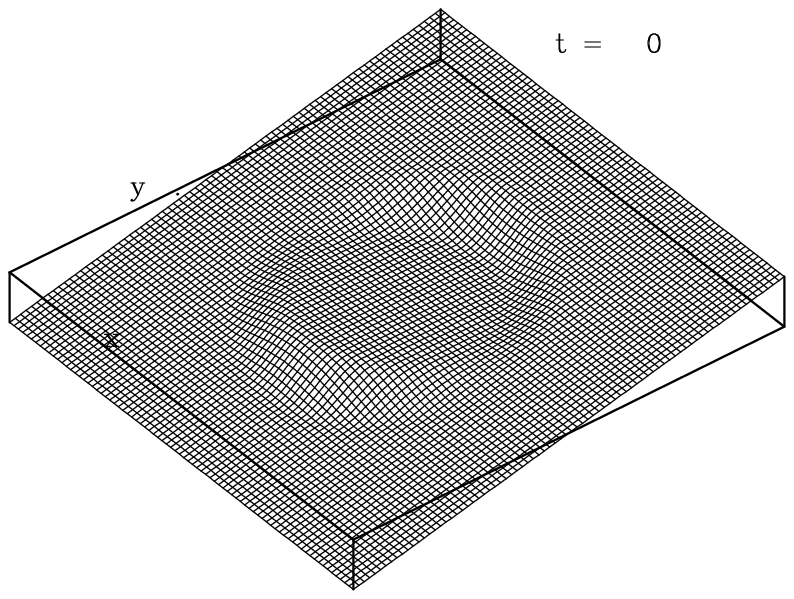}
\includegraphics[width=.32\hsize]{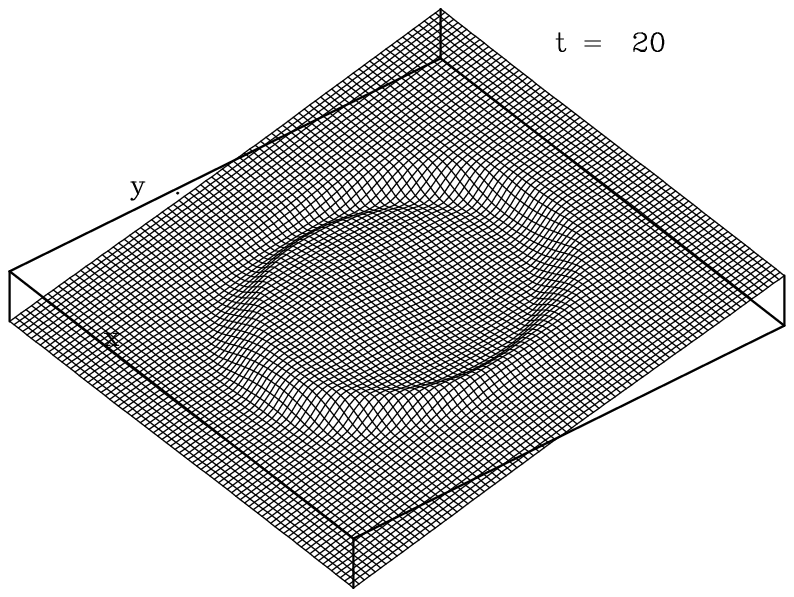}
\includegraphics[width=.32\hsize]{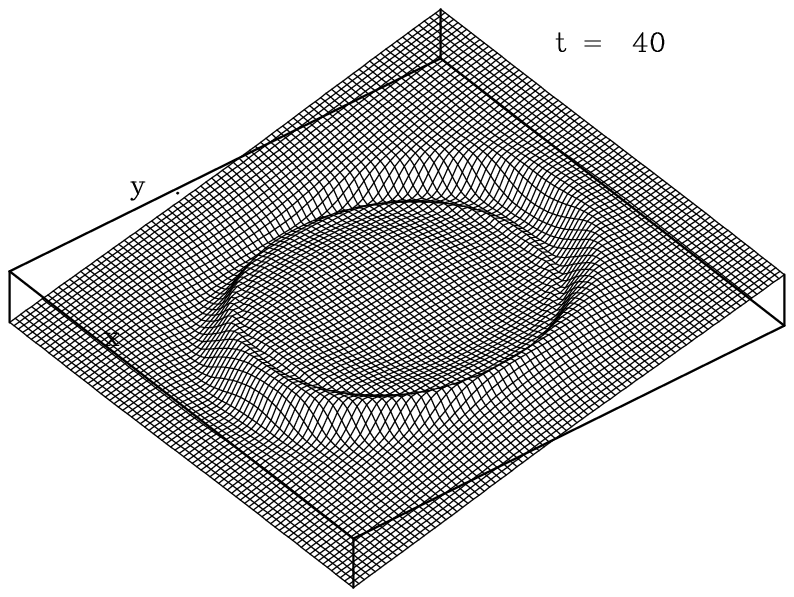}}
\centerline{
\includegraphics[width=.32\hsize]{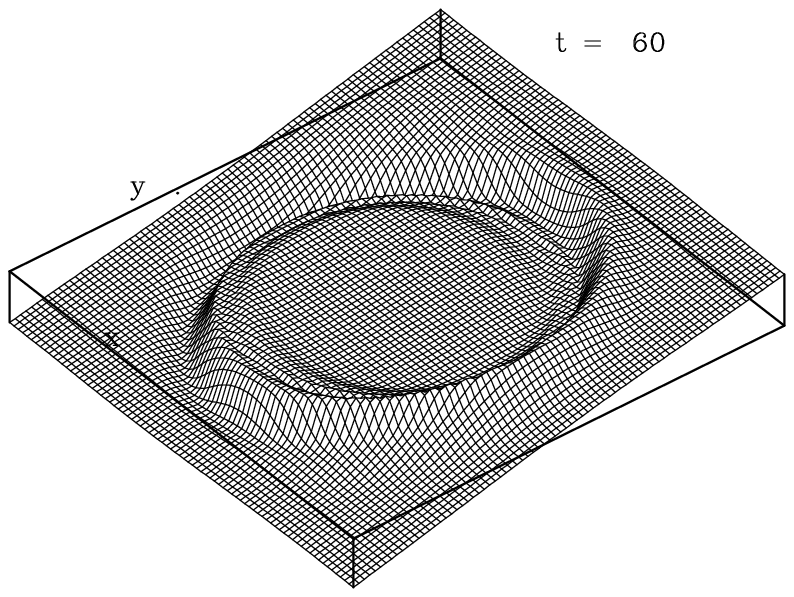}
\includegraphics[width=.32\hsize]{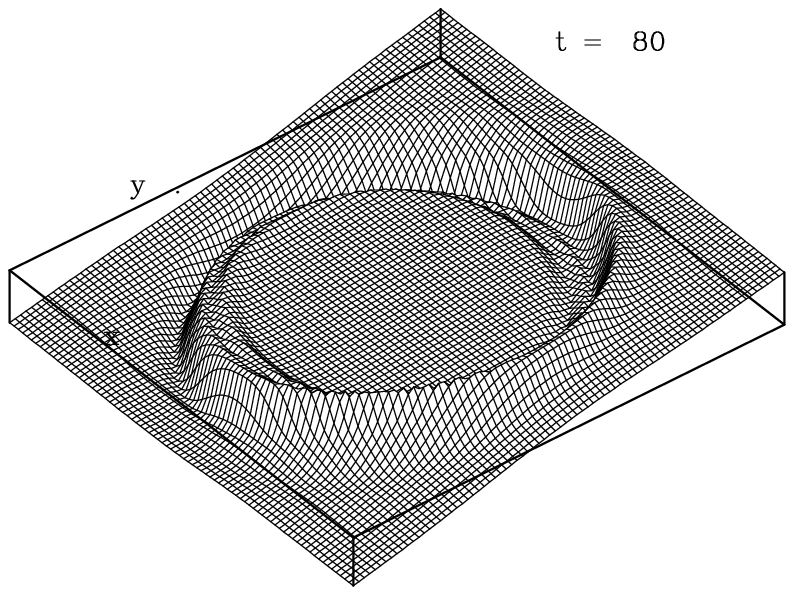}
\includegraphics[width=.32\hsize]{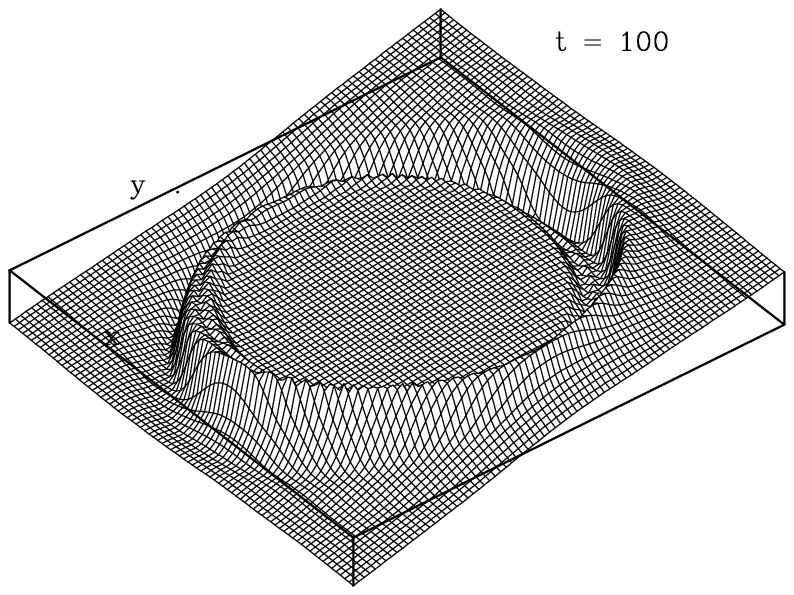}}
\centerline{
\includegraphics[width=.32\hsize]{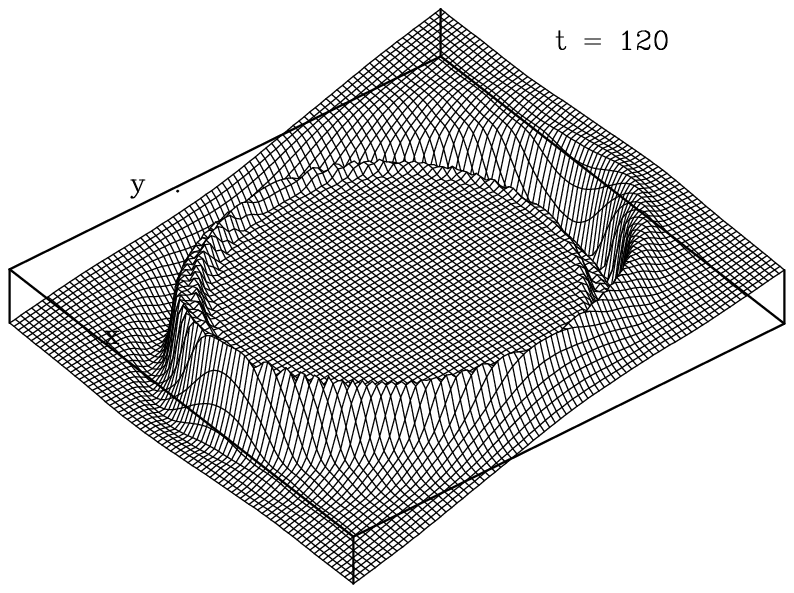}
\includegraphics[width=.32\hsize]{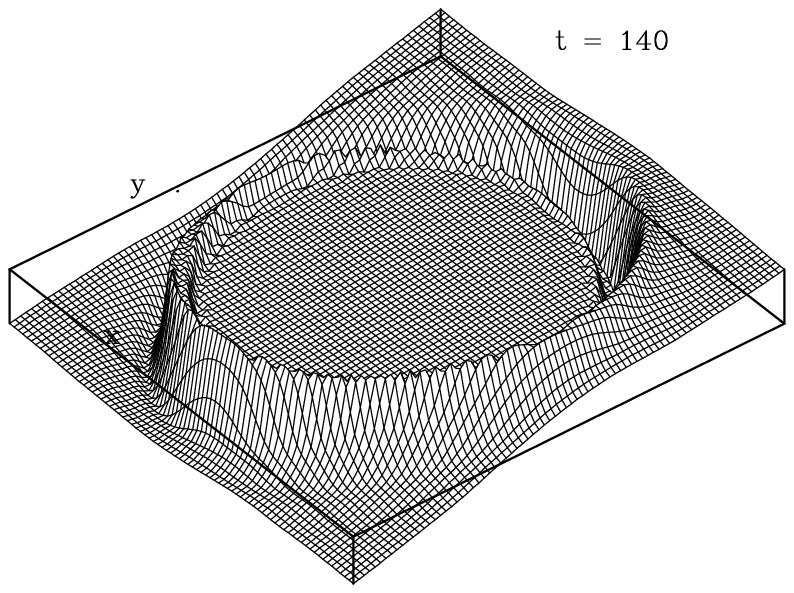}
\includegraphics[width=.32\hsize]{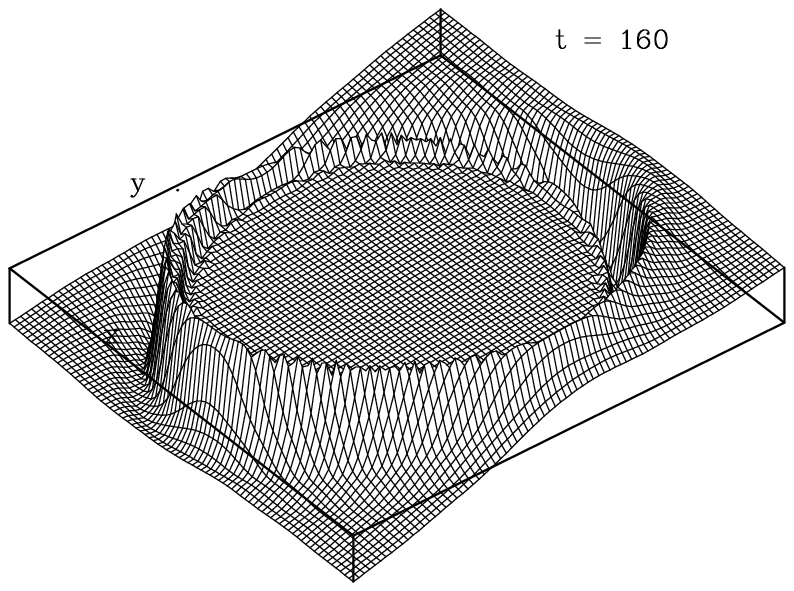}}
\caption{An $m=1$ disturbance in an exponential disc at the times given above in units
of $\sqrt{R_\d/G\Sigma_0}$. The plotted grid extends to $x,y=\pm7.1R_\d$ while
the radial grid used for the computation extends to
$R=10R_\d$. The \rms\ variation of energy during the integration was $2.7$
per cent.}\label{fig:freeWarp}
\end{figure*}

\section{Warping of an isolated Galaxy}\label{sec:freeWarp}

In this section we integrate the equations from a non-zero initial
disturbance but without external stimulus such as that from the Sgr dwarf.
These integrations both elucidate the physics of warps and enable us to check
on the correctness of the code by evaluating the energy integral
\eqref{eq:finalHTintegral} {\rd that is derived in Appendix \ref{sec:appE}}. We focus on the dipole component $h_1$, when
stars populated rigid rings.

The dotted curve in Fig.~\ref{fig:freeMWVc} shows the circular speed generated by
an exponential disc with an
exponential surface-density profile
\[
\Sigma(R)=\Sigma_0\e^{-R/R_\d},
\]
 while the dashed curve shows the circular speed of the dark halo adopted in
his section. The
combined circular-speed curve is not dissimilar to that of our own
Galaxy. Notice that the orbital period at $R=3R_\d$ is $T_\phi=2\pi
R/v_c\simeq3\pi\sqrt{R_\d/G\Sigma_0}$.

Fig.~\ref{fig:freeWarp} shows the evolution of a disc that initially lies in
different planes at small and large radii: at $R<2.5R_\d$ the disc lies in
$z=0$ while at $R>4.5R_\d$ it lies in a plane that intersects $z=0$ along the
$x$ axis:
\[
\Im h_1(R)=-h_{\rm max}\times\begin{cases}
0&R<\overline R -\Delta\cr
R/( \overline R+\Delta)&R>\overline R+\Delta\cr
\fracj12\Big(1+\sin\Big[{\pi\over2\Delta}(R-\overline R)\Big]\Big)&
\hbox{otherwise}.
\end{cases}
\]
with $\overline R=3.5R_\d$ and $\Delta=R_\d$.
Here $h_{\rm max}$ is an arbitrary constant that plays no role because we
are doing linear theory.

\begin{figure}
\centerline{\includegraphics[width=.8\hsize]{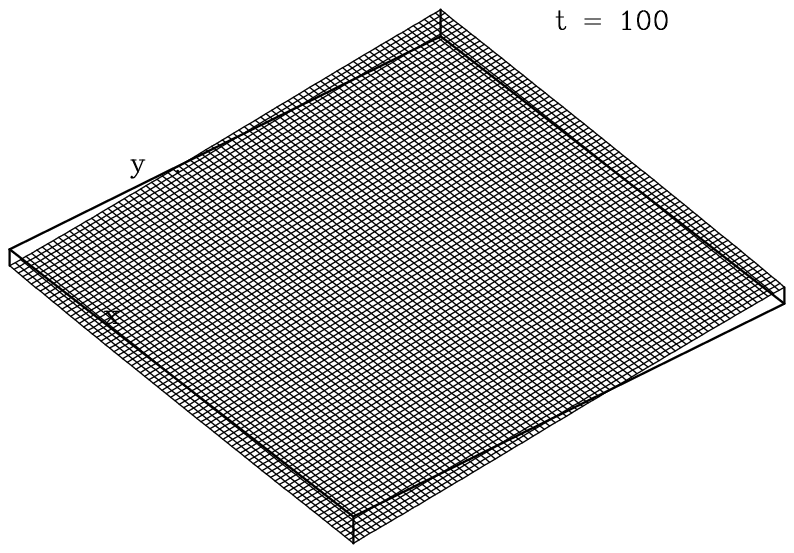}}
\caption{The inner disc at $t=100$ is to good precision planar but tilted so
the line of nodes makes and angle of $20\,$deg with the $x$ axis. The largest
ring contained within the square has $R=2.5R_\d$.}\label{fig:T100}
\end{figure}

The top row of panels in Fig.~\ref{fig:freeWarp} shows the evolution of the
disc over the first 160 dynamical times $\sqrt{R_\d/G\Sigma_0}$.
The flat outer region torques the flat inner region about the $x$ axis, while
the inner region torques the outer region in the opposite sense about the
same axis.  Both regions respond to torques like gyroscopes by tilting around
the $y$ axis: in Fig.~\ref{fig:freeWarp} the near part of the inner region
moves downwards, while the furthest part moves upwards (Fig.~\ref{fig:T100}).
The rings that surround the inner disc move in the opposite direction and
within the outer disc the line of nodes tilts away from the $x$ axis in the
opposite sense to the line of nodes across the inner disc.

The near edge of the outer region moves downwards.  The
torques are strongest at the inner and outer edges of the transition region
in which $h_1$ varies rapidly. The coupling between rings decays
exponentially outwards with the surface density, so the inner region's rings
remain almost coplanar by efficiently redistributing angular momentum that's
acquired from the outer region. The outer rings are much less tightly
coupled, with the result that the angular momentum acquired from the inner
disc remains concentrated in the rings within and just outside the transition
region. In consequence these rings precess (retrogradely) rather rapidly. In
Fig.~\ref{fig:freeWarp} the tipping and precession of these rings manifests
as a bump on the near side and a trench on the far side.

\begin{figure}
\centerline{\includegraphics[width=\hsize]{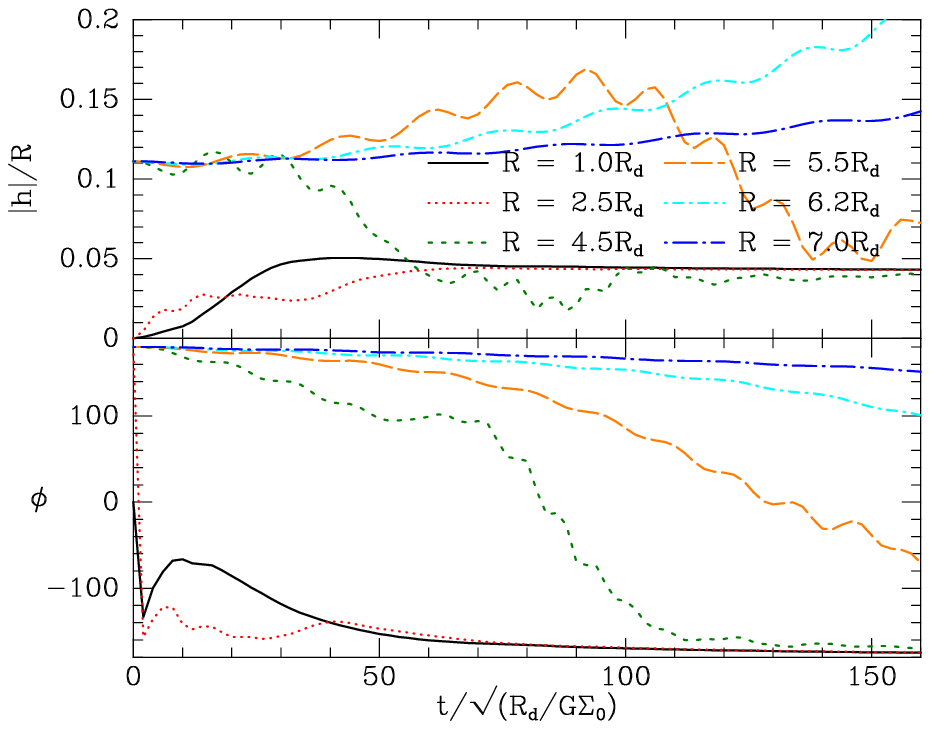}}
\caption{The warp's amplitude (upper panel) and line of nodes (lower panel) as
functions of time on six selected rings.}\label{fig:freeWtimes}
\end{figure}

A better understanding of the evolution seen in Fig.~\ref{fig:freeWarp} can
be gained by examining plots of $|h_1|/R$ and the angle $\phi$ of the line of
nodes as functions of time for representative rings.
Fig.~\ref{fig:freeWtimes} plots these variables for six rings: two in the
central region ($R=R_\d$ and $R=2R_\d$), two that the transition region
passes over ($R=4.5R_\d$ and $R=5.5R_\d$) and two in the outer region
($R=6.2R_\d$ and $R=7R_\d$).

The upper panel of Fig.~\ref{fig:freeWtimes} shows that $|h_1|/R$ is
initially constant over the outer three and rises only slowly later. This
indicates that the tilts of these rings are consistent with them
continuing to lie in the same inclined plane. By contrast, the two rings of
the central region have $|h_1|/R$ initially rising from zero and then
flattening out to a common constant, signalling that from $t\sim60$ onwards
the inner region settles to a flat disc that's inclined with respect to the
coordinate system.  The curve of $|h_1|/R$ for the third ring from the centre
($R=4.5R_\d$) starts from the value characteristic of the outer rings but
from $t\sim30$ falls towards the steady value reached by the inner two rings.
This behaviour reflects the spreading of the central, flat region that is
evident in Fig.~\ref{fig:freeWarp}.

The lower panel of Fig.~\ref{fig:freeWtimes} shows the position angle of the
lines of nodes  from equation (\ref{eq:lons}) for each of the six rings. The curves for the outer four rings
start at $180\,$deg and fall at rates that decrease as $R$ increases because a
ring's precession rate decreases with $R$. The two innermost rings initially
have undefined values of $\phi$ because they lie in the plane $z=0$, so their
curves in Fig.~\ref{fig:freeWtimes}
are initially erratic. By $t\sim10$, the full curve for the innermost ring
has settled to value around $-70\,$deg and from there it falls steadily
towards $\phi\sim-160\,$deg. The curve for the next ring out, $R=2.5R_\d$,
converges on the curve for $R=R_\d$ by $t\simeq40$.
Thus the inner disc tilts but ultimately does not precess. The
curve for the ring with $R=4.5R_\d$ falls rapidly until $t\sim120$, and then  oscillates around
$\phi=-155\,$deg, reflecting this ring's absorption by the flat inner disc.

\begin{figure}
\centerline{\includegraphics[width=\hsize]{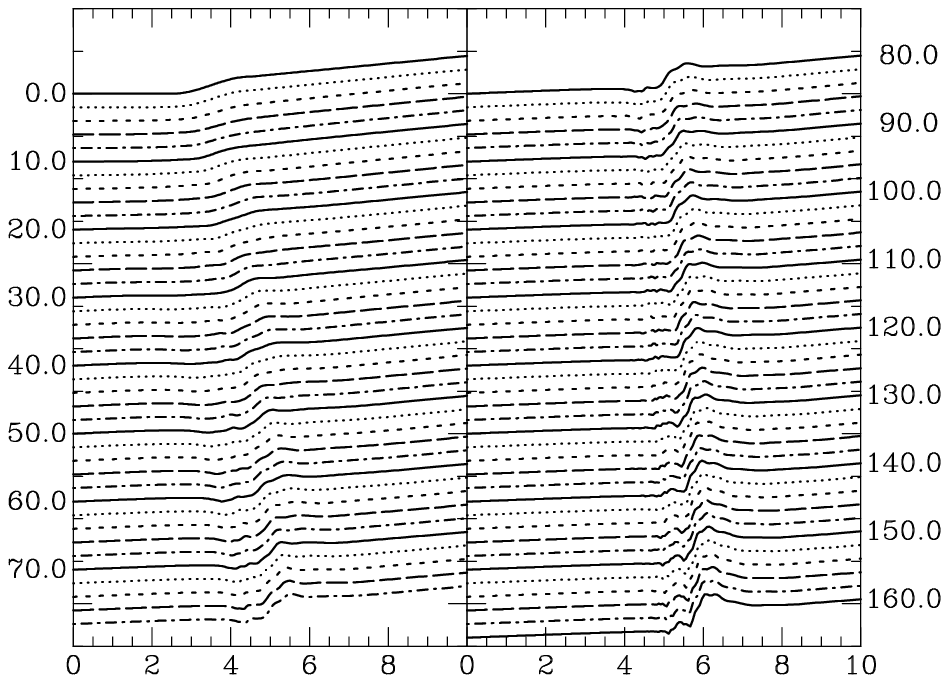}}
 \caption{The amplitude of a corrugation with $m=1$
within a razor-thin exponential disc. Each curve shows the amplitude $|h_1|$
as a function of $R$ {\rd in kpc} at a given time; successive curves are displaced
downwards for clarity. The numbers down the left and right edges give the
times in units of $\sqrt{R_\d/G\Sigma_0}$ of the nearest full curve.}
\label{fig:bendwaveone}
\end{figure}

Fig.~\ref{fig:bendwaveone} provides a simpler perspective on the disc's
evolution by plotting the amplitude $|h_1|$ as a function of radius at a
large number of times. Each curve has a well-defined step in the annulus
within which the inner, flat disc transitions to the outer tilted disc.  This
region of transition moves steadily outwards -- over 160 time units ($\sim
5$ orbital periods at $4R_\d$) its centre moves from $3.5R_\d$ to
$5.75R_\d$. Inside the transition region, a small linear increase in $|h_1|$
develops, because the inner disc tilts slightly while remaining almost flat.
Outside the transition region, the initial linear increase of $|h_1|$ with
$R$ fades between $80$ and $160$ time units.  The length of the transition
region does not change much.

\begin{figure}
\centerline{\includegraphics[width=\hsize]{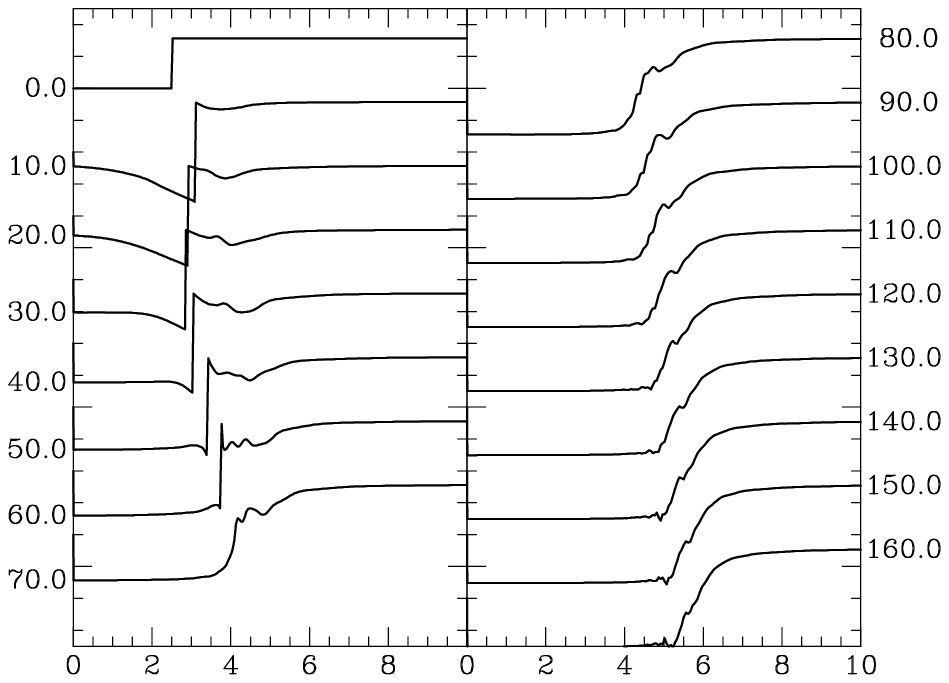}}
\caption{The angle $\phi$ in the $xy$ plane of the line of nodes from
equation (\ref{eq:lons}) as a function of ring radius {\rd in kpc}. The times of the full curves are
given along the left and right edges as in Fig.~\ref{fig:bendwaveone}.}\label{fig:bendwavetwo}
\end{figure}

Fig.~\ref{fig:bendwavetwo} complements Fig.~\ref{fig:bendwaveone} by plotting
the position angle $\phi(R)$ of the line of nodes as a function of radius at
a series of times. At $t=10$ $\phi$ varies quite rapidly in the inner disc,
implying that the disc has a spiral structure. From $t\sim60$ torques have
straightened the line of nodes in the inner disc. 

Before $t\sim70$, there are steps in $\phi(R)$ that are caused by $\phi$
moving through $\pm180\,$deg. They can be banished by
redefining  the definition of $\phi$, but with $\phi$ redefined, the curves
acquire step discontinuities at $t>60$. Regardless of how $\phi$ is defined,
the orientation of the line of nodes swings rapidly from one side of the $x$
axis to the other as $R$ passes through the transition zone.

\begin{figure}
\centerline{\includegraphics[width=.8\hsize]{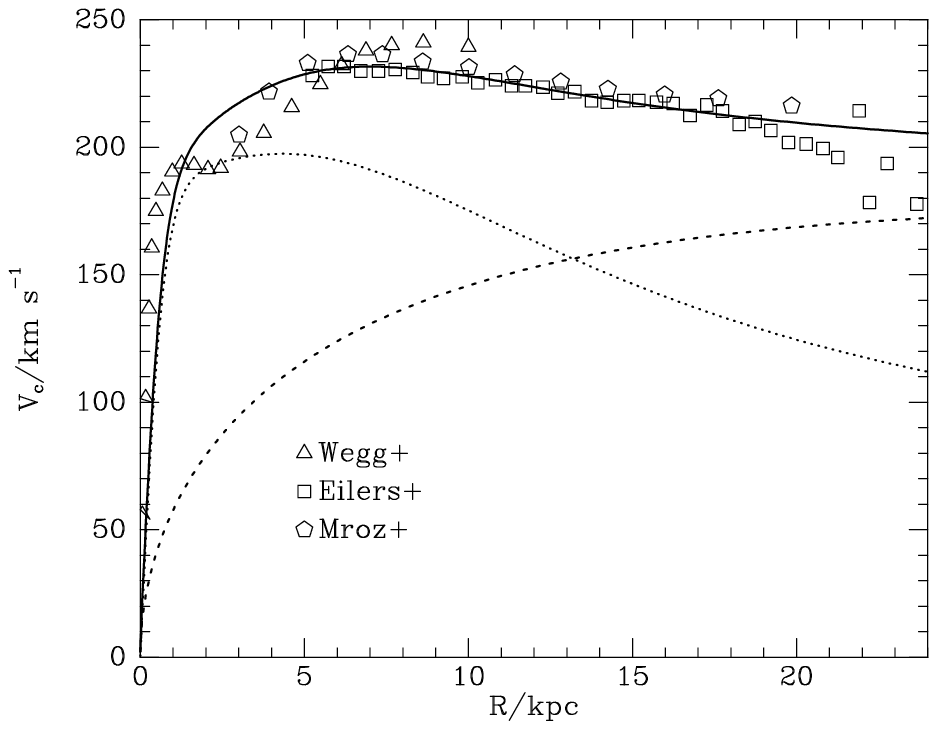}}
\caption{Full curve: the circular speed in the adopted Galaxy model. Dotted
and dashed curves show the contributions of baryons and dark matter,
respectively. The points show observational estimates from three studies.}\label{fig:MWcirc}
\end{figure}

\section{Warping by the Sgr dwarf}\label{sec:SgrWarp}

\subsection{Orbit of the Sgr darf}
Possible orbits of the Sgr dwarf have been explored by \cite{DierickxLoeb2017},
\cite{VasilievBE2021} and many others.  Table~\ref{tab:Sgr} gives the current
phase space coordinates of the Sgr dwarf from \cite{VasilievBE2021}. We have
integrated orbits of point masses in the Galaxy's potential to find one that
comes close to this location. For simplicity, we assume the Galaxy's
potential is spherical, with contributions from a \cite{He90}
sphere, a disc and a bulge. At the radii explored by the Sgr dwarf, the
potentials of the disc and bulge can be approximated by those of point masses
and we have used this approximation.  The mass and scale length of the
Hernquist sphere are chosen such that, after addition of the (not necessarily
spherical) potentials of the
model disc and bulge listed in Table~\ref{tab:DiscBulge}, we obtain the fit to
the Galaxy's circular-speed curve that is shown in Fig.~\ref{fig:MWcirc}. 

\begin{table}
\caption{The current phase-space location of the Sgr dwarf.}\label{tab:Sgr}
\begin{tabular}{cccc}
$(\alpha,\delta)/$deg & $s/$kpc & $(\mu_\ell,\mu_b)/$mas\,yr$^{-1}$ & $V_{\rm
los}/\!\kms$ \\
\hline
$(283.76,-30.48)$ & $27$ & $(-2.7,-1.35)$  & $142$\\
\end{tabular}
\end{table}

The assumption of spherical symmetry confines the dwarf's orbit to an orbital
plane. We compute orbits in the presence of dynamical friction,
computed from equation (8.3) in \cite{GDII} using the sphere's ergodic DF
(eqn.~4.51 in Binney \& Tremaine). 

\begin{figure}
\centerline{\includegraphics[width=.8\hsize]{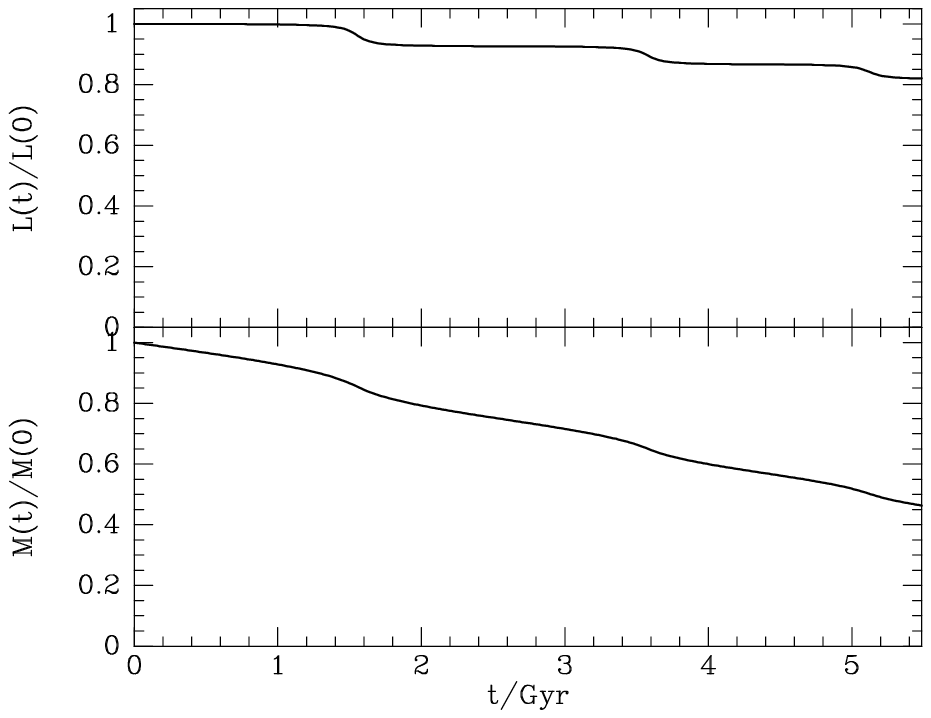}}
\caption{The dwarf's angular momentum (upper
panel) and mass (lower panel) along the orbit shown in Fig.~\ref{fig:SgrXZ}.}\label{fig:SgrMJ}
\end{figure}

\begin{table}
\caption{Models of the Galaxy's disc and bulge }\label{tab:DiscBulge}
\begin{tabular}{lccc}
Component & Type & mass & scale length\\
&& $10^{10}\msun$ &kpc\\
\hline
Disc&exponential&5&3\\
Bukge&Plummer sphere&1.5&1\\
Dark halo&Hernquist sphere&100&35\\
\end{tabular}
\end{table}

\begin{figure*}% decay.cpp
\centerline{\hfil\includegraphics[width=.4\hsize]{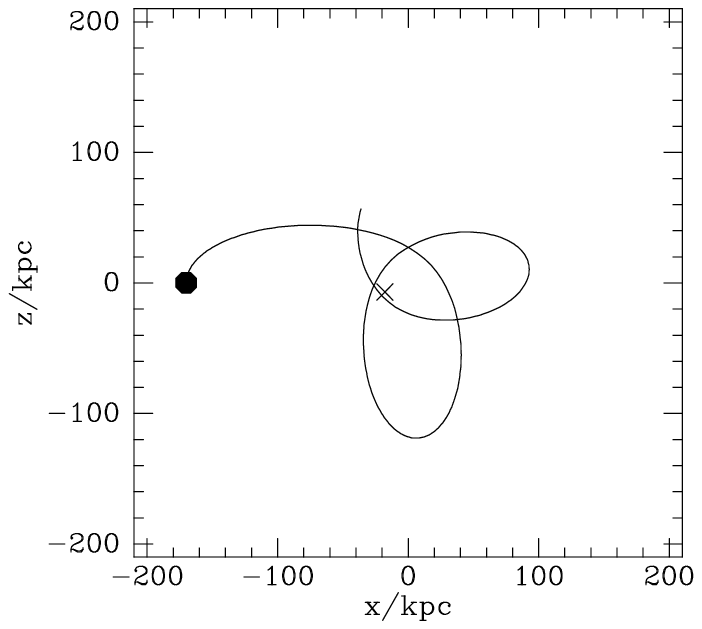}\hfil
\includegraphics[width=.4\hsize]{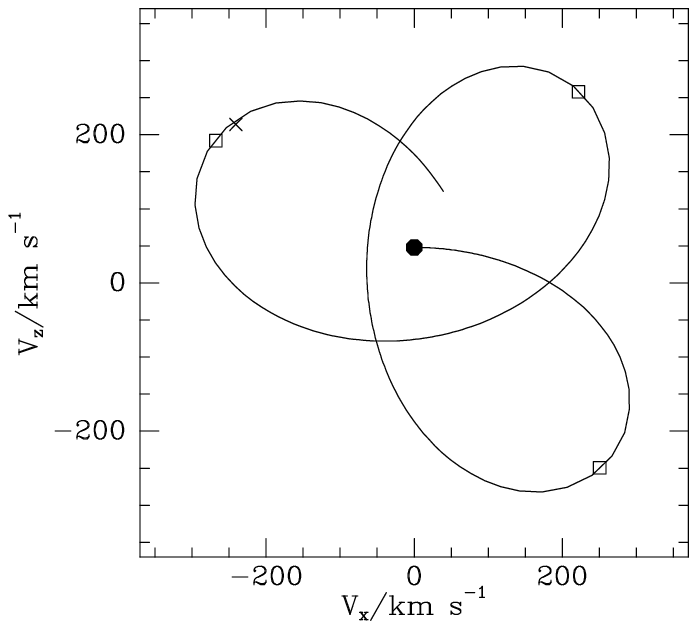}\hfil}
\caption{A possible orbit of the Sgr dwarf in real space (left) and velocity
space (right). The black dots show the initial conditions chosen, while the
crosses sow the observed location of the dwarf. Pericentres are marked by
open squares.}\label{fig:SgrXZ}
\end{figure*}

\begin{figure}% plotTide.cpp using data from decay.cpp
\centerline{\includegraphics[width=.8\hsize]{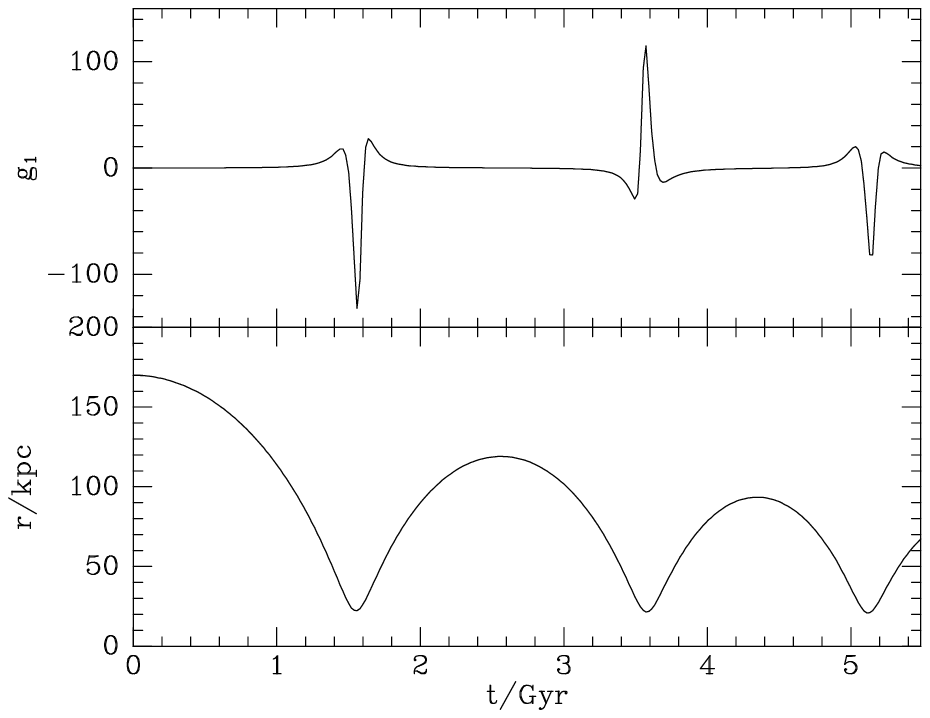}}
\caption{Galactocentric radius (lower panel) and dipole forcing at
$R=12\kpc$ (upper
panel) as functions of time
along the orbit shown in Fig.~\ref{fig:SgrXZ}. {\rd The units of $g_1$ are
$\msun\pc^{-2}$.} }\label{fig:plotTide}
\end{figure}

\begin{figure}
\centerline{\includegraphics[width=.9\hsize]{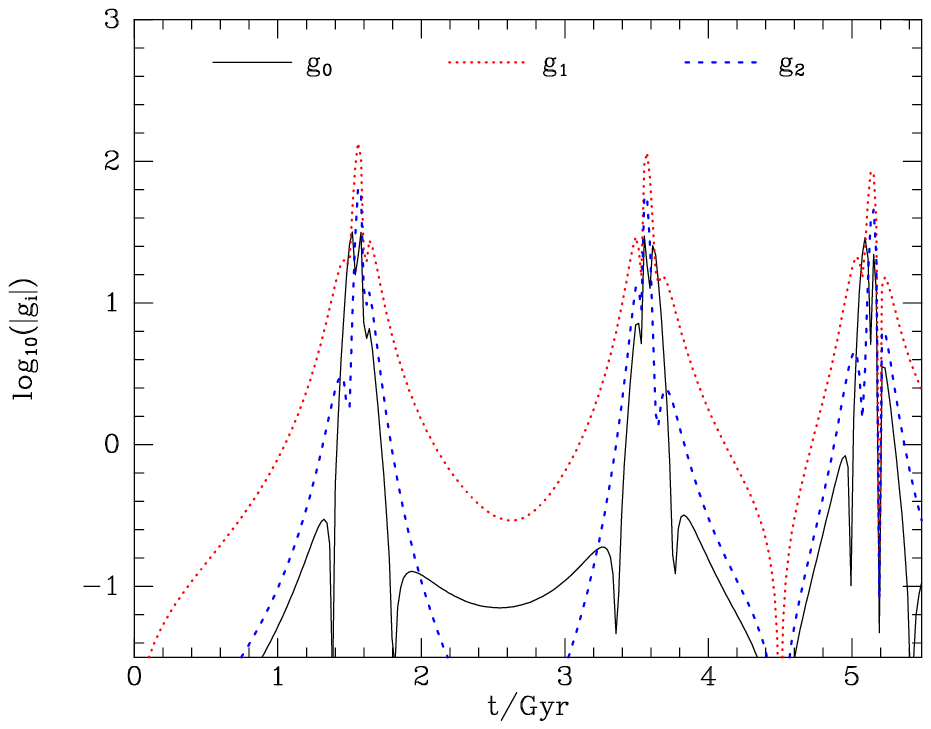}}
\caption{The magnitudes of the monopole (full curve), dipole (dotted curve)
and quadrupole (dashed curve) contributions
to tidal forcing of the Galactic disc at $T=12\kpc$ by the Sgr
dwarf. {\rd The units of $g_m$ are $\msun\pc^{-2}$.}}\label{fig:plotTidetwo}
\end{figure}

Tidal stripping of the dwarf's mass was computed by modelling the dwarf as a
Hernquist sphere of mass $m$ and scale radius $a$. At each timestep,
the dwarf's tidal radius $r_{\rm t}$ was computed from the equation 
\[\label{eq:tidal}
{m\over(r_{\rm t}+a)^2r_{\rm t}}={M\over(R+A)^2R},
\]
 where $M$ and $A$ are the mass and radius of the Galaxy's halo and $R$ is
the distance between the centres of the dwarf and the Galaxy. Equation
(\ref{eq:tidal}), which can be expressed as a cubic that typically has only one
real root, equates the acceleration at radius $r_{\rm t}$ generated by the dwarf
to the difference in the forces applied by the Galaxy at $r_{\rm t}$ and at the dwarf's
centre. Let $m(t)$ be the dwarf's mass at time $t$ and let $M_{\rm t}$ be
the mass within $r_{\rm t}$ of a Hernquist sphere of total mass $m$, and
scale radius $a=5\kpc$, i.e.,
\[
M_{\rm t}=m(t)\Big({r_{\rm t}\over r_{\rm t}+a}\Big)^2.
\]
Then we have evolved $m$ by solving
\[
{\d m\over\d t}=-{m-M_{\rm t}\over T_{\rm strip}},
\]
 where $T_{\rm strip}$ is a constant of order a gigayear. This procedure
acknowledges that material that lies outside $r_{\rm t}$ does not instantly
vanish, while material that at time $t$ lies inside the tidal radius will
eventually be stripped. The lower plot of Fig.~\ref{fig:SgrMJ} shows the how
$m$ decreases with time along our favoured orbit, which is obtained with
$T_{\rm strip}=2\Gyr$. The decline steepens slightly at each pericentre --
corresponding but clearer features are visible in the plot of angular
momentum versus time in the upper panel of Fig.~\ref{fig:SgrMJ}.

Fig.~\ref{fig:SgrXZ} shows our orbit in real space on the left and in
velocity space on the right. Black dots mark initial conditions and crosses
mark the observed location of the Sgr dwarf; squares mark pericentres.
$35\Myr$
after its third pericentre and $5.156\Gyr$ since the start of the integration,
the orbit passes close to the dwarf's observed location. This orbit was
obtained by setting $m(0)=5\times10^{10}\msun$ and taking the  Coulomb
logarithm to be $\ln\Lambda=6$, but the orbit depends only on the product
$m(0)\ln\Lambda$. 

\begin{table}
\caption{Pericentres along the dwarf's orbit. The dwarf reaches its current
location at $t=5.156\Gyr$.}\label{tab:Orbit}
\begin{center}
\begin{tabular}{lccc}
&Peri 1&Peri 2&Peri 3\\
\hline
$r$ (kpc)&22.32&21.54&20.92\\
$t$ (Gyr)&1.552&3.578&5.121\\
\end{tabular}
\end{center}
\end{table}

\begin{figure*}%plot_MWgas.cpp controlled by SgrWarp.in
\centerline{
\includegraphics[width=.4\hsize]{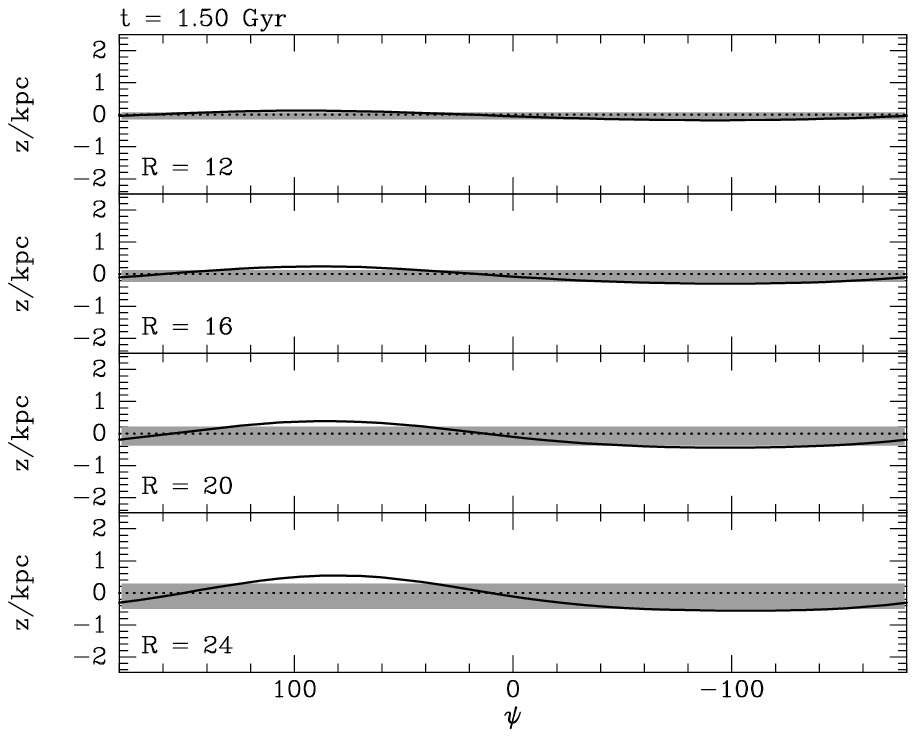}\qquad
\includegraphics[width=.4\hsize]{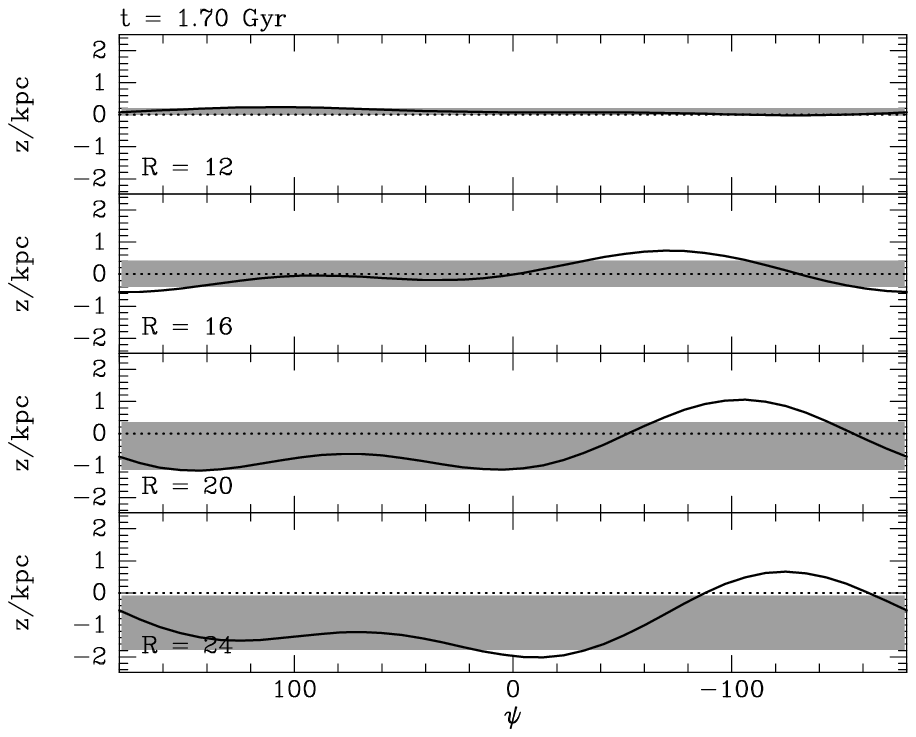}
}
\centerline{
\includegraphics[width=.4\hsize]{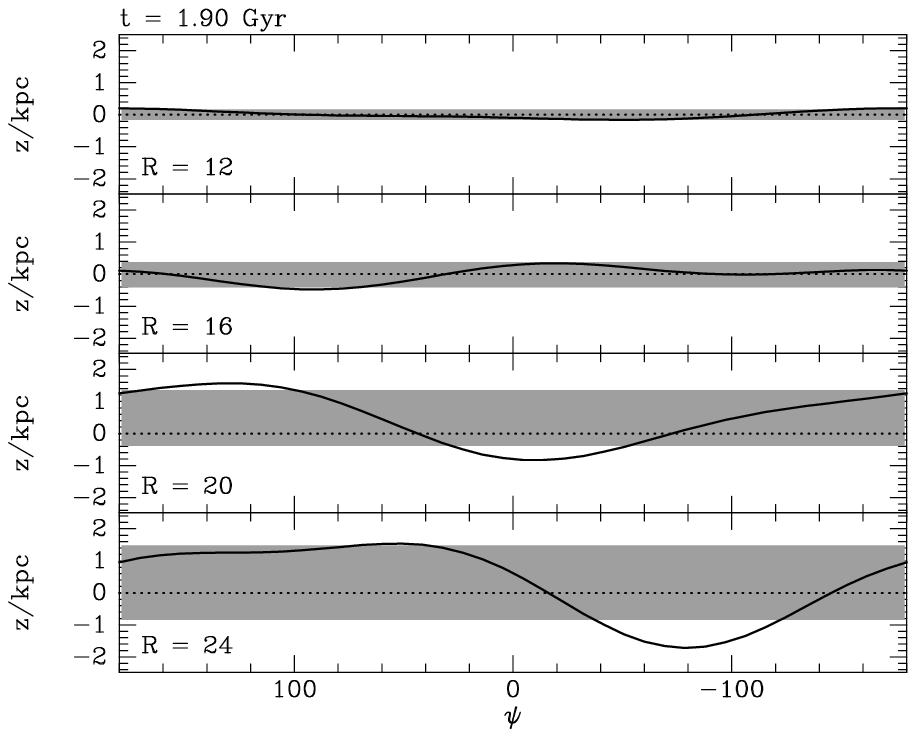}\qquad
\includegraphics[width=.4\hsize]{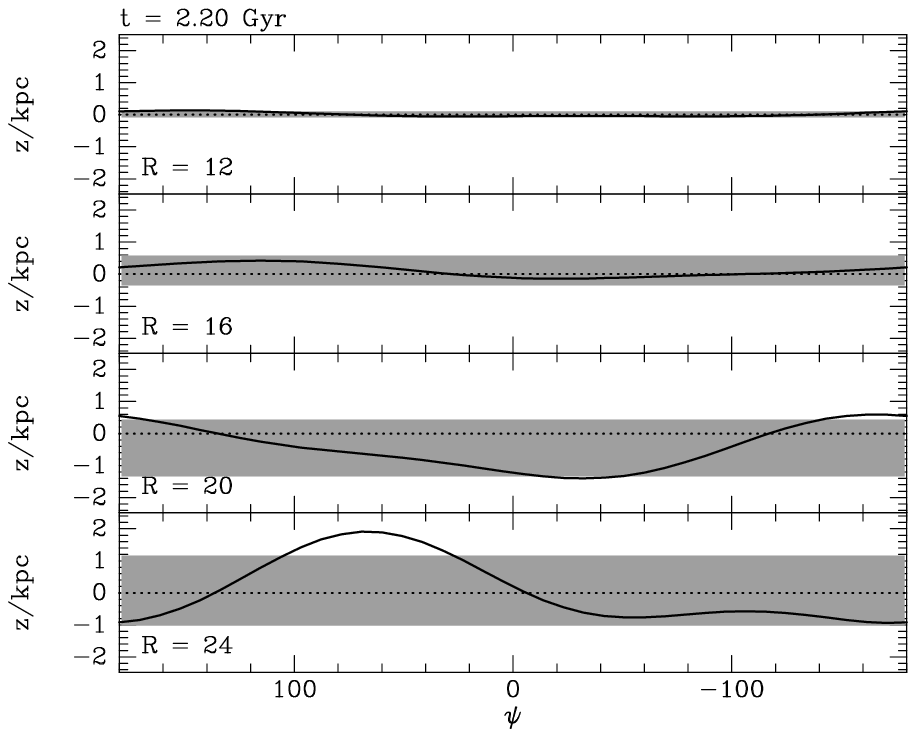}
}
\caption{Amounts $z(\psi)$ by which stars are pulled from the plane averaged
over  four annuli centred on  $R=12,16,20$ and $24\kpc$, containing 24, 18,
15 and
12 rings, respectively. The curves are plotted from the values of $h_0$,
$h_1$ and $h_2$ radially averaged with weight $\e^{-R/5\kpc}$, while the
grey bands show the  corresponding standard deviations of $z$. The times,
given at top left of each panel, span the dwarf's first
pericentre. The angle $\psi$ increases in the direction of Galactic
rotation.}\label{fig:FirstTotal} 
\end{figure*}

\begin{figure}%plot_MWgas.cpp controlled by SgrWarp.in
\centerline{\includegraphics[width=.87\hsize]{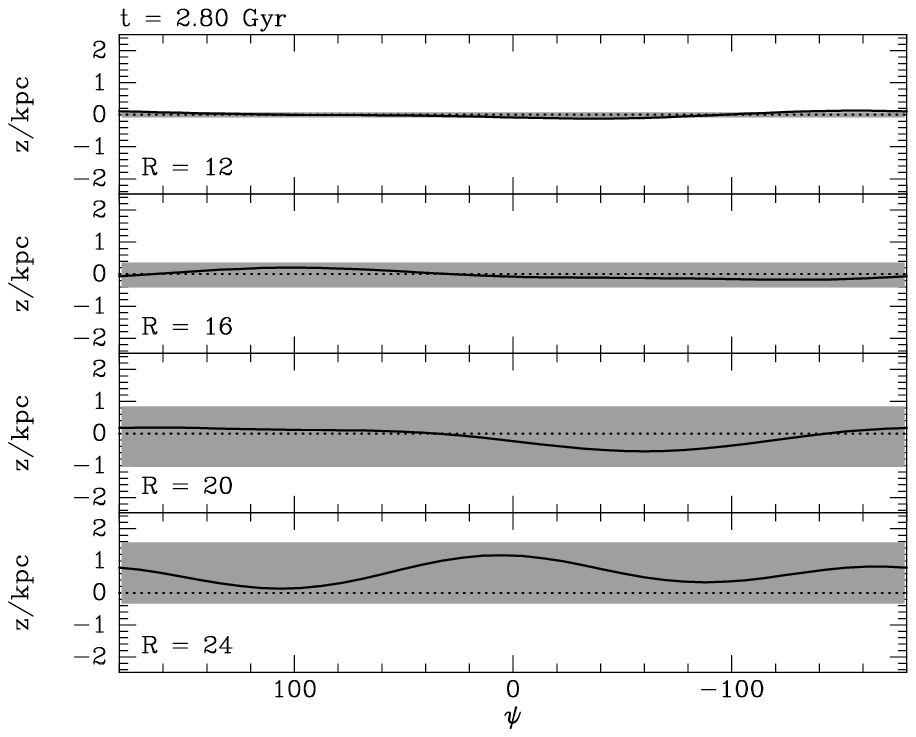}}
\caption{The distortion of the disc at $t=2.8\Gyr$, shortly after the
dwarf's second apocentre. The meaning of the curves and bands is as in
Fig.~\ref{fig:FirstTotal}.}\label{fig:TotalRest}
\end{figure}

\begin{figure*}
\centerline{
\includegraphics[width=.3\hsize]{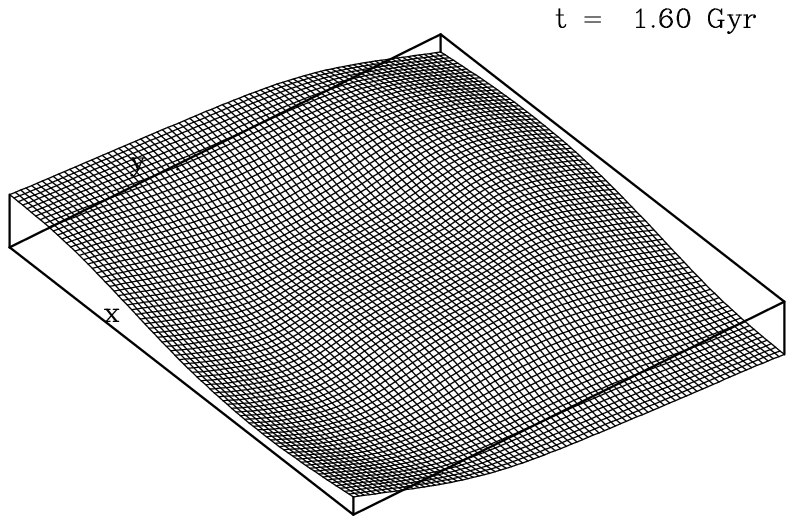}\ 
\includegraphics[width=.3\hsize]{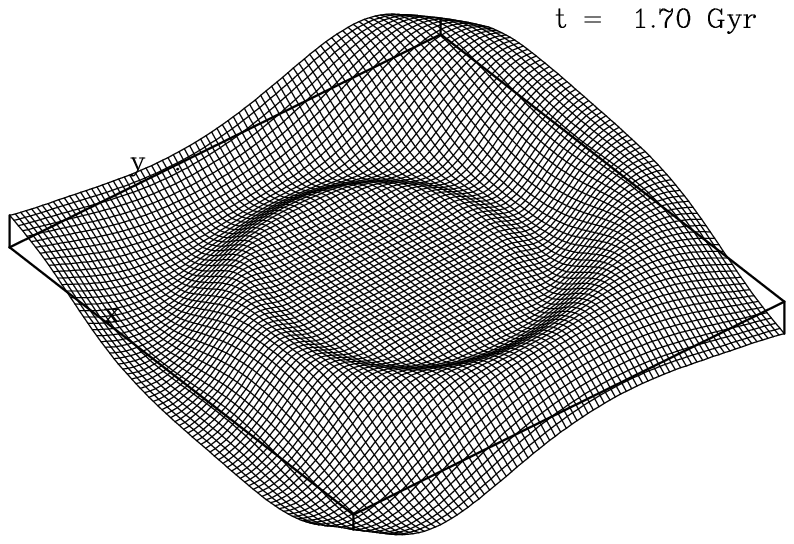}\ 
\includegraphics[width=.3\hsize]{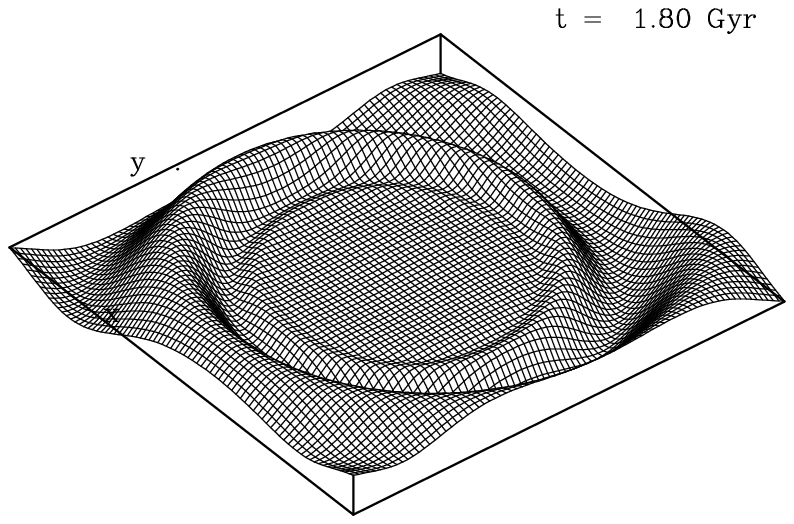} 
}
\centerline{
\includegraphics[width=.3\hsize]{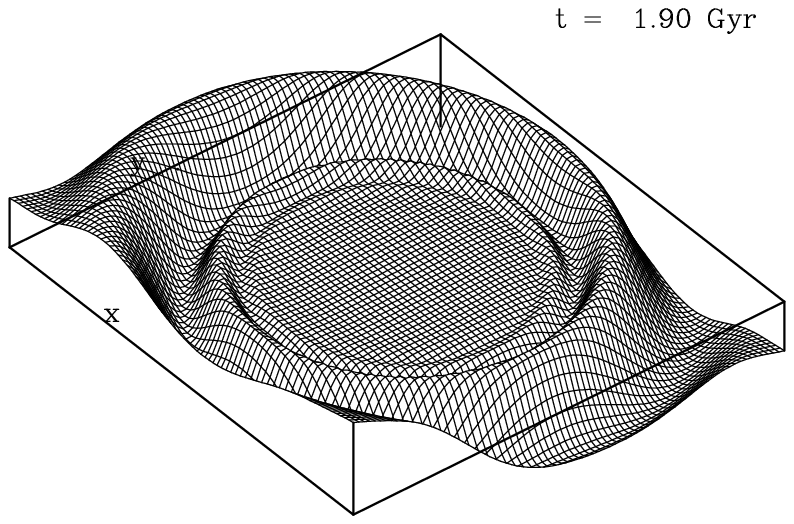}\ 
\includegraphics[width=.3\hsize]{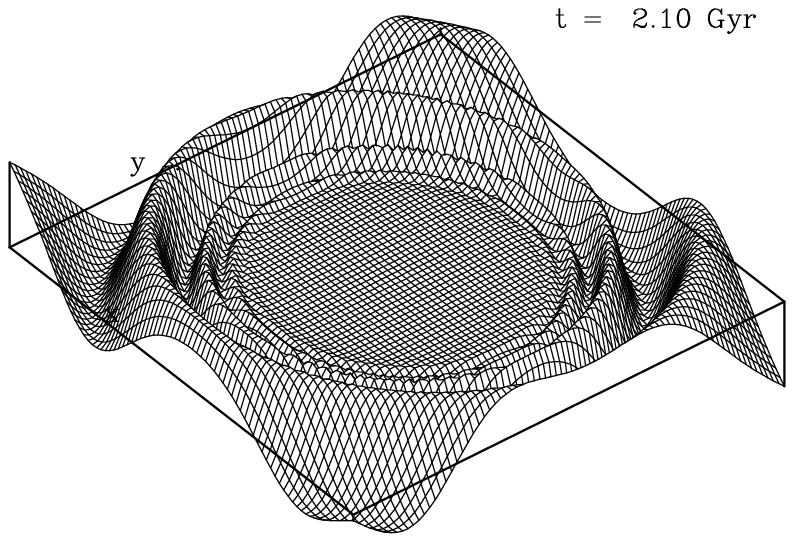}\ 
\includegraphics[width=.3\hsize]{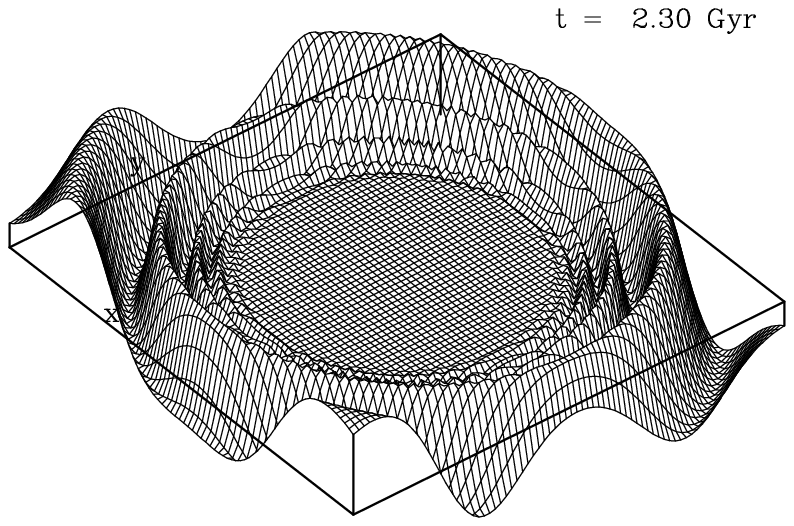}
}
\caption{The component $m=1$ of the distortion that is induced in the
Galactic disc by the Sgr dwarf
galaxy at its first pericentre. The squares are $42.4\kpc$ on a side so they
completely contain rings out to radius $21.2\kpc$.}\label{fig:Sgr3d}
\end{figure*}

\begin{figure}% plot_zsds.cpp
\centerline{\includegraphics[width=.8\hsize]{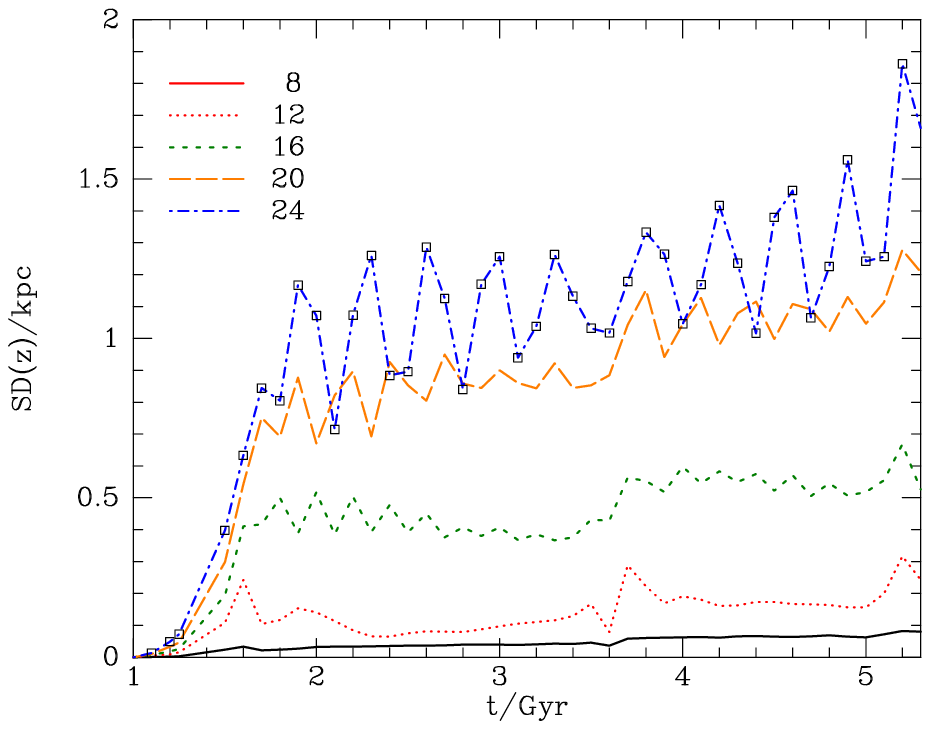}}
\caption{Each curve shows the evolution of the standard deviation of $z$ in
an annulus $2\kpc$
wide as the dwarf's orbit is integrated from $t=1\Gyr$ to $t=5.3\Gyr$, which
is about $150\Myr$ into the future. The central radius of each annulus is shown at top
left. In the case of the curve for $R=24\kpc$, times at which $\Delta_z$ has
been calculated are marked by squares. These times apply to all curves.}\label{fig:plot_zsds}
\end{figure}
\begin{figure}
\centerline{\includegraphics[width=.7\hsize]{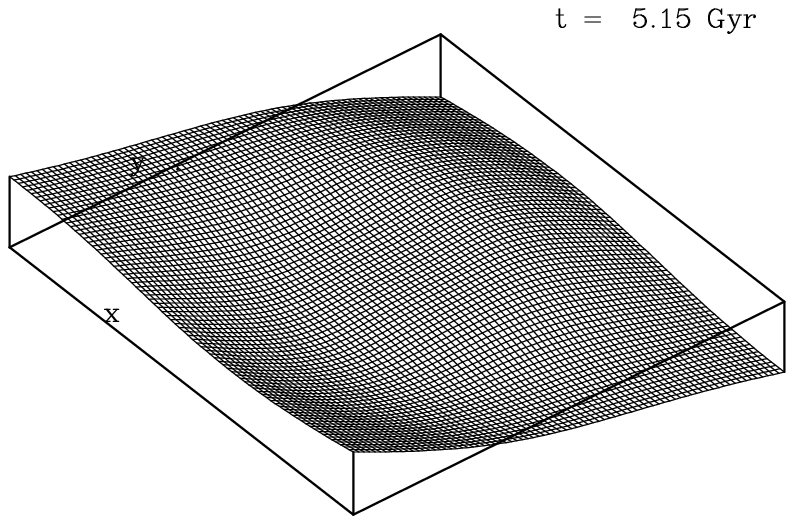}}
\centerline{\includegraphics[width=.7\hsize]{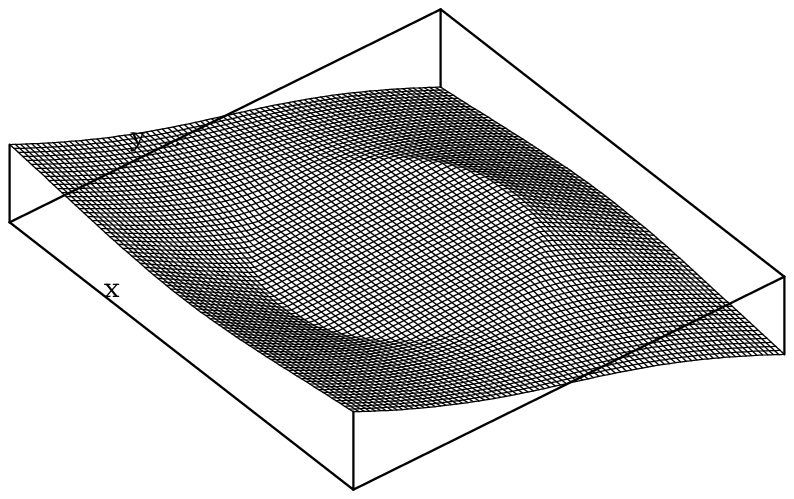}}
\caption{Upper panel: the $m=1$ structure of the disc at $t=5.15\Gyr$ after an initially flat
disc has been evolved from $t=4.6\Gyr$ through the most recent pericentre of
the Sgr dwarf. Lower panel: the $m=1$ component of the model fitted to HI
data by Binney \& Merrifield (1998).}\label{fig:DiscNow}
\end{figure}

The orbit is computed as a planar object in the approximation that the
Galactic potential is spherical, and is then rotated into the frame that's
aligned with the Galactic disc and places the Sun on the negative $x$ axis.
At any given time the tidal acceleration $\vg$ can now be computed from
equation (\ref{eq:tidalg}) and Fourier decomposed (eq.~\ref{eq:tidalFT}).
The lower panel of Fig.~\ref{fig:plotTide} shows the Galactocentric radius of
the dwarf as a function of time.  The pericentric radii, given in
Table~\ref{tab:Orbit}, decrease remarkably slowly. The near constancy of the
pericentric radii reflects much faster loss to friction of energy than of
angular momentum. This in turn reflects the density profile of the Galaxy's
dark halo: since a tangential force $F$ causes energy and angular momentum to
change at rates $\dot E=vF$ and $\dot L=rF$,
at peri- or apocentre, the rates of loss of $E$ or $L$ are in the
ratio
\[
{\dot E\over\dot L}={v\over r},
\]
where $v$ is the dwarf's speed and $r$ is its Galactocentric radius. The
ratio $v/r$ is much bigger at pericentre than at apocentre, so drag at
pericentre tends to circularise the orbit, while drag at apocentre has the
opposite tendency. Our use of a Hernquist dark halo, which has an asymptotic
variation of density $\rho\propto r^{-3}$ rather than that $\rho\propto
r^{-2}$ characteristic of an NFW density profile \citep{NFW97} favours
circularisation. We have, however, checked that computing friction from and
NFW profile has a negligible effect on the evolution of the pericentres,
probably because the dwarf's orbit spans the scale radius $a=35\kpc$ around
which the slope of the logarithmic density profile transitions from $-1$. 

The upper panel of Fig.~\ref{fig:plotTide} shows the coefficient
of $\cos\phi$ in the dipolar component  of the tidal acceleration on the
ring of radius $R=12\kpc$ -- the coefficient of $\sin\phi$ is
significantly smaller. We see that the forcing is very tightly concentrated
around pericentres. Fig.~\ref{fig:plotTidetwo} reveals that around each
pericentre the monopole and quadrupole contributions to tidal forcing are not
much smaller than the dipole contribution. This finding may explain why the
distortion of our Galaxy's HI distribution requires significant contributions
from monopole and
quadrupole terms \citep[cf][eq.~9.18]{GA}.

\subsection{Warp excited by the dwarf}

Since the Galaxy rotates clockwise when viewed from the north, in the
equations of motion (\ref{eq:warpeqs}) we take
$\Omega<0$. Within the Galactic plane we use polar coordinates $(R,\psi)$
with $\psi$ increasing in the direction of Galactic rotation.

Fig.~\ref{fig:FirstTotal} shows the amounts $z(\psi)$ by which the stars are
pulled from the plane at four times around the dwarf's first pericentre: the
curves show the values of $z$ obtained by averaging the Fourier coefficients
$h_i$ for $i\le2$ over annuli $3\kpc$ wide that are centred on $R=12,16,20$
and $24\kpc$. The grey bands show the standard deviations of $z$ within these
annuli, which are readily calculated from the mean values of $h_i$ and
$|h_i|^2$.  The distortions are very small until the dwarf reaches pericentre
at $t\sim1.6\Gyr$. Then it grows rapidly at $R\ga10\kpc$, and by $1.7\Gyr$ it
reaches $2\kpc$ at $R\sim24\kpc$. The subsequent development comprises a
train of waves that moves outwards whilst shifting to shorter wavelengths. As
the dwarf moves away, the radially-averaged amplitude of the vertical
oscillations slowly diminishes at any given radius through phase mixing and
the outward transport of energy by waves.  Fig.~\ref{fig:TotalRest}
illustrates this result by showing the mean and standard deviation of
$z(\psi)$ within the annuli at $2.8\Gyr$, that is shortly after the
dwarf's second apocentre. The curves show smaller amplitudes of oscillation
than in the bottom-right panel of Fig.~\ref{fig:FirstTotal} while the widths
of the grey bands have changed very little.

 Fig.~\ref{fig:Sgr3d} shows a three-dimensional representation of the disc at
six times spanning $0.7\Gyr$ from about the first pericentre. The highest point
moves clockwise by almost a quarter turn in the $0.1\Gyr$ that separates the
first four snapshots and has completed a more than a complete turn by the
fifth snapshot at $t=2.1\Gyr$. Near the centre we see a flat region that
slowly expands and is bounded by crests and troughs that form a two-armed
trailing spiral that becomes rapidly tighter. The overall effect is to create
a stadium in which a flat playing field is surrounded by tiers of an
ever-increasing number of benches that become steadily narrower.

The pericentre that provokes the response plotted in Fig.~\ref{fig:Sgr3d}
lies $\sim3.5\Gyr$ in the past, so the winding up process that is already
advanced in the final snapshot would by now have brought peaks and troughs so
close to one another that the in-plane epicycle motion of stars that is
absent from the model would have completely blended peaks and troughs into a
smooth, warp-free disc that {\rd has become thick outside  the central flat
region. In other words, the long-term effect of the first pericentre is to
make an initially thin disc into one that flares strongly at the edge of the
central flat region.}
Fig.~\ref{fig:plot_zsds} shows as a function of time the standard deviation
$\Delta_z$ of $z(R,\phi)$ within annuli that are $2\kpc$ wide and centred on
$R=8,12,16,20$ and $24\kpc$.  We should interpret $\Delta_z$ as the disc
thicknesses that result from the winding up of the warp that the dwarf
induces as it orbits through its last three pericentres. We see that at
$R=24\kpc$ $\Delta_z$ shoots up just before the dwarf passes through its first
pericentre at $t\sim1.5\Gyr$. By $t=2\Gyr$ the sharp upward movement of
$\Delta_z$ has flattened to a gradual upward drift to $\sim1.5\kpc$ on which
are superposed oscillations that have a period $\sim0.4\Gyr$. The behaviour
of $\Delta_z$ at $R=20$ and $16\kpc$ is similar although the level reached at
$t=2\Gyr$ is smaller, especially at $R=16\kpc$, and the later oscillations
have smaller amplitude and a shorter period. For thee radii it's just
possible to detect upwards steps in $\Delta_z(t)$ around the times, $3.5\Gyr$
and $5.15\Gyr$, of the dwarf's second and third pericentres. At $R=8\kpc$ and
$5\kpc$ (not plotted) $\Delta_z$ remains less than $0.1\kpc$ throughout. At
$12\kpc$ $\Delta_z$ is currently $\sim0.2\kpc$, which is significantly less
than the half thickness of the low-$\alpha$ disc near the Sun.
  
Figs.~\ref{fig:FirstTotal} to \ref{fig:plot_zsds} thus lead to the conclusion
that soon after a pericentre, a warp develops that co-rotates with the stars
and winds up from inside out.  In the long run the pericentre produces an
axisymmetric but flaring disc.  For only $\sim0.3-0.4\Gyr$ from pericentre
will a warp will be evident.  In light of this conclusion we now focus on the
effect that the most recent pericentre has on the gas disc, {\rd because we expect
viscous damping to re-establish a thin gas disc after each pericentre.}
Moreover the configuration of the gas disc far beyond the solar circle is
better constrained by observations than is the configuration of the outer
reaches of the stellar disc.

The upper panel of Fig.~\ref{fig:DiscNow} shows the disc as it would be now
if it had been flat $\sim0.95\Gyr$ ago as the Sgr dwarf moved past its last
apocentre. The following pericentre would have generated a warp that
depresses the disc in the direction of Galactic rotation i.e., right back in
Fig.~\ref{fig:DiscNow}. The lower panel of Fig.~\ref{fig:DiscNow} shows the
$m=1$ component of the fit to observations of the Galaxy's HI disc given in
\cite{GA}. At a qualitative level the agreement between model and fit to data
is good: model and fit agree on the extent of the disc's flat portion, and
that the line of nodes of the large-scale warp falls near the $x$ axis (on
which the Sun lies). The figure conceals a significant difference, however,
by using scales for $z$ that differ by a factor three: the amplitude of the
observed warp is three times larger than that of the model warp. There are
several possible explanations of this discrepancy, as we discuss below.

\begin{figure}
\centerline{\includegraphics[width=.85\hsize]{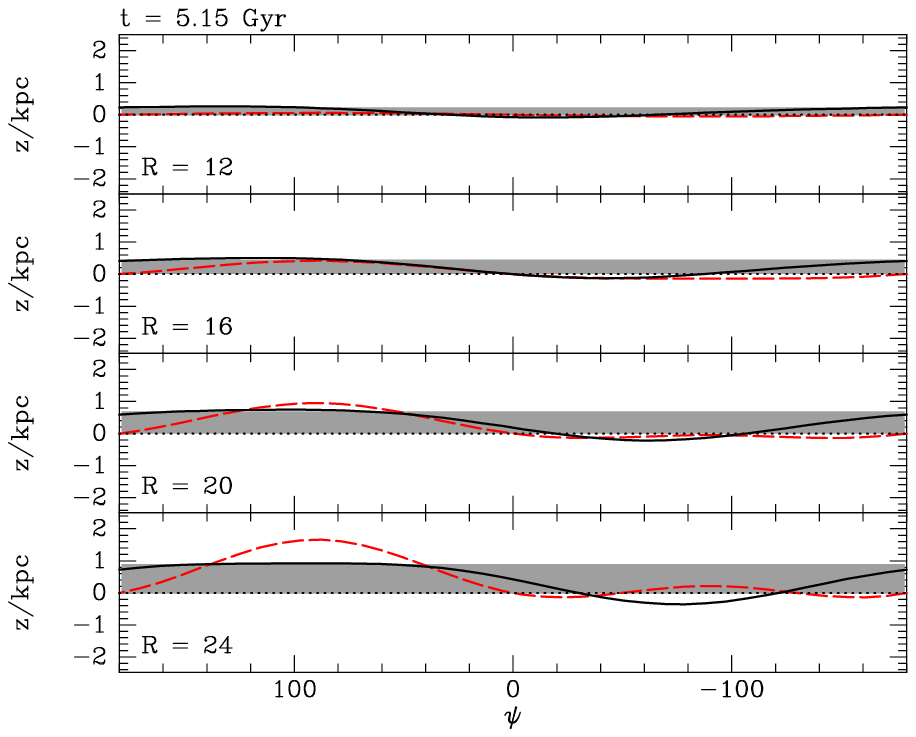}}
\caption{Plots of $z(\psi)$ at four radii at the current time according to
the model of the effect of the Sgr dwarf. The dashed red curves show
$z(\psi)/3$ where $z(\psi)$ is  the fit
of equation (9.18) in Binney \& Merrifield (1998) to the model of the
Galactic HI layer depicted in their Figure 9.24. }\label{fig:BinneyMerrifield}
\end{figure}

Fig.~\ref{fig:BinneyMerrifield} plots the model's prediction for the location
$z(\psi)$ of the middle of the gas layer around three annuli, $R=12,16,20$ and
$24\kpc$. These plots may be compared with Figure 9.24 in \cite{GA}, which
shows the estimated density $n(z,\psi)$ of HI within cylindrical
shells.\footnote{The caption to Figure 9.24 is in error: the Sun lies at
$\psi=\pm180\,$ deg, not at 0.} The model captures the asymmetry between
positive and negative values of $\psi$, which is defined so it increases in
the direction of Galactic rotation and the sun lies on $\psi=\pm180\,$deg.
The model also captures much of the growth with $R$ of the amplitude of the
warp. 

When the equations are integrated into the future, the warp is seen to change
rapidly: the central region becomes flatter and more sharply demarcated from
the outer region, and the latter becomes corrugated rather than warped. 

\section{Discussion}\label{sec:discuss}

The vertical dynamics of galactic discs are a potentially fruitful area, not
least for the light they could shine on density fluctuations in dark haloes.
Unfortunately, this is an exceptionally challenging area of stellar dynamics
because the vertical oscillations of stars are strongly anharmonic and
inherently coupled to their in-plane motions. Moreover, we suffer from a want
of sufficiently accurate analytical expressions for the gravitational
potential of a distorted disc of finite thickness.  Consequently our options
seem to be (i) brute-force N-body simulation
\citep[e.g.][]{Laporte2018,JBH_Galah2019,JBHThor2020}, which is expensive and
cumbersome, (ii) evolving non-equilibrium distribution functions in a fixed model
potential $\Phi(z)$ \citep[e.g.][]{BinneySchoenrich2018,TremaineFB2023}, which is
very restrictive, and (ii) as here, modelling a disc by stars on perfectly
circular orbits, which excludes any discussion of the distributions of stars
in the $(z,v_z)$ phase plane.

This phase plane has recently attracted a lot attention following the
discovery by \cite{AntojaSpiral} of a spiral in the $(z,v_z)$ distribution of
stars near the Sun. Unfortunately, it's logically impossible to deduce the
impact of a perturber such as the dwarf on the $(z,v_z)$ distribution of
stars without computing the bodily acceleration of the disc at
each radius by the combined gravitational pulls of the perturber, the rest of
the disc and the dark halo. The focus of the present paper has been precisely
on this larger-scale dynamics which should logically precede discussion of
dynamics within $(z,v_z)$ phase planes.

Our discussion of the evolution of a free warp in Section~\ref{sec:freeWarp}
was instructive but limited in scope by the initial conditions having $\dot
h=0$. It is likely that other, possibly more interesting, behaviour will
emerge when $\dot h$ is chosen differently. In the case of a violin string,
setting $\dot h$ to zero is equivalent to specifying initial conditions that
comprise two waves that differ only in their direction of travel. {\rd Our
observation in Section~\ref{sec:freeWarp} that the disturbance seems to move
outwards could be considered to confound this expectation. A disturbance {\it
does} move inwards, however, but it tips the inner disc in an orderly fashion
rather than generating tightly-wound corrugation waves.}

Probably the most important uncertainties in Section~\ref{sec:SgrWarp}
concern the mass and shape of the Sgr dwarf. At early times the dwarf's mass
has to be of order $10^{11}\msun$ if it is to experience enough dynamical
friction to get onto its present orbit \citep{JiangBinney}. We have rather
arbitrarily adopted $5\times10^{10}\msun$ as initial mass and our algorithm
for mass loss causes the mass to fall to $1\times10^{10}\msun$, which is
quite a large value given the dwarf's current luminosity $L\sim10^8\lsun$
\citep{NiedersteOstholt2012}. Clearly,
changes in the dwarf's mass will translate immediately to changes in the
amplitude of the warp it generates.

An equally serious uncertainty is the location of the dwarf's mass. A
significant fraction of the dwarf's stars have already been stretched into a
stream that wraps more than once around the Galaxy. The dwarf's dark matter
is even more liable to shredding into a stream and we ought really to be
computing the response of the disc to a stream that spreads while moving its
centre along the computed orbit. 

We have compared our dynamical model to observations by way of a fit to a
model of the HI distribution. The HI model was fitted by
\cite{VoskesBurton1999} to a composite data cube extracted from the surveys
of \cite{HartmannBurton1997}, \cite{WeWi74} and \cite{Keea86}. The model
specifies the density $n(z,\psi)$ of neutral hydrogen in a series of annuli.
Extracting such a model from the observational data cubes is not
straightforward and Voskes \& Burton never published a full account of their
work -- the fit we have used is based on four plots that Voskes and Burton
provided for publication in \cite{GA}. The model relies on distances to
emitting hydrogen that assume that {\rd the Sun's radius} $R_0=8.5\kpc$ and
the circular speed is constant at $220\kms$. It follows that the HI model,
and its analytic fit do not have the status of fact.

A dynamical model would be better compared with HI data by using the model to
generate a mock HI data cube. The drawback of this procedure is the
difficulty of displaying the quality of fits in a three-dimensional space,
but this procedure should nonetheless be tried soon.

A major issue is that the dynamical model produces an amplitude that is three
times smaller than observations seem to require, even though it is based on a
model of the dwarf that is surely too concentrated and would be expected to
over- rather than under-estimate its effect on the disc. The model has,
however, two shortcomings that
likely lead to under-estimation of the disc's response. 

\begin{itemize}

\item
One is failure to
model the dynamical response of the dark halo, which we have assumed to be
spherical rather than flattened. The halo must respond strongly to the dwarf
-- and the model exploits this response in as much as it gives rise to
dynamical friction -- but our equations of motion (\ref{eq:warpeqs}) completely
ignore the halo's response. \cite{Weinberg1989} explains how to compute the
wake and shows that its structure is materially modified when the wake's
self-gravity is taken into account. It's worth noting that self-gravity {\it
reduces} the frictional drag to which the wake gives rise, by reducing the
offset between the wake's barycentre and the perturber's location.
Consequently, our treatment either over-estimates the frictional drag or
under-estimates the dwarf's mass. 

\item Another questionable feature of the model is its assumption that the HI
is part of a razor-thin disc with a strictly exponential surface density that
is extrapolated from the {\it stellar} surface density near the solar circle.
We are certain that the dwarf's latest pericentre is not its first, and we
have argued that the first pericentre caused the disc to flair strongly from
$R\sim4R_\d$. After this first pericentre, the model applies only to the gas
disc, which has a much lower surface density than the model assumes. On
account of its low surface density, the gas disc is {\rd much less strongly
stiffened by gravity} than
the model assumes, and will consequently respond more strongly to the current
pericentre.

\end{itemize}

\section{Conclusions}\label{sec:conclude}

Section~\ref{sec:eqs} offers a derivation of the equations of motion for the
vertical dynamics of a razor-thin disc that differs in significant respects
from that given by HT69. The principal difference is the use here of standard
cylindrical polar coordinates rather than the confocal ellipsoidal system
employed in HT69, but the equations derived are not mere transcriptions of
the old equations between coordinate systems. The infinitesimal thickness of
the disc gives rise to subtle questions about limits when one generalises
from a series of discrete rings to a continuous disc. We have formulated
equations which yield finite values for the quantities required for the
equations of motion even in the continuum limit.

The change of coordinate system frees the disc from the
requirement to lie within the infinitely-flattened ellipsoid of the confocal
system, and therefore make it possible to discuss the dynamics of a standard
exponential disc. The disc's potential can be computed in this system quite
cheaply thanks to the Green's function that \cite{CohlTohline1999}
introduced.

Section~\ref{sec:freeWarp} illustrated the tendency of warps to wind up into
corrugation waves that propagate outwards, leaving the inner, higher-density,
part of the disc flat. In Section~\ref{sec:SgrWarp} we presented a highly
idealised model of the interaction of the Sgr dwarf with the Galactic disc.
The tidal field of the dwarf is strongly concentrated around the pericentres
of the dwarf's orbit. The current phase-space coordinates of the dwarf imply
that the dwarf has passed at least three pericentres, and the radii of
successive pericentres decrease extremely slowly because, for any plausible
structure of the Galaxy's dark halo, the dwarf's orbit circularises. From
this it follows that the decisive pericentre must have been the first one, at
least $3.6\Gyr$ ago. It seems that then the stellar disc beyond $\sim4R_\d$ must
have been thickened as the corrugation waves generated by the dwarf wound up.
This conclusion chimes with accumulating evidence that near the solar circle
the `thick disc' transitions from being a high-$\alpha$ structure to a
low-$\alpha$ structure \citep{SchusterNissen2012,BinneyVasiliev2023}. An
alternative, and probably better, perspective on the data is that just
outside solar circle, the low-$\alpha$ disc suddenly flares into a thick structure
just as our model requires.

In light of the  likely disruption of the outer thin disc $\ga3.6\Gyr$ ago,
our razor-thin model disc can be applied  to the dwarf's latest pericentre
only by treating it as a model of the HI layer, which we expect to re-form
into a thin disc after each disruption.

Considering the crudeness of the model and the number of poorly constrained
parameters involved, the extent of the agreement between theory and
observation in Figs.~\ref{fig:DiscNow} and \ref{fig:BinneyMerrifield} is encouraging.
The amplitude of the response is a factor three smaller than the HI data
imply but an under-estimated amplitude is to be expected in view of
the model's assumption of an unrealistically high surface density for the gas
disc and a rigid halo. What is gratifying is the extent to which the model captures the
complex shape of the warp, which has significant $m=0$ and $m=2$ components in
addition to the dominant dipole component. Moreover, the warp's shape
evolves remarkably quickly and it agrees with the observed shape rather
fleetingly. It's striking that the moment of agreement coincides precisely
with the moment that the dwarf passes through its  observed phase-space
location.

This investigation suggests several avenues for further investigation.
\begin{itemize}
\item Add the Magellanic Cloud system to the problem. This would be simple if
the Cloud system were modelled in the same way as the dwarf, namely as a
Hernquist sphere of mass $\sim10^{11}\msun$ that orbits a spherical model of
the Galaxy as if it were a point particle. \cite{WeinbergBlitz2006} long ago
argued that when tidal distortion of the dark halo is taken into account, the
Cloud system  {\rd acting alone} can account for the outer gas warp. Recently it has
been argued that the Cloud system significantly perturbs the orbits of
objects such as globular clusters that orbit in the distant halo
\citep{Erkal2019}. Hence it is likely that they have a non-negligible effect
on the disc at $R\ga20\kpc$ even if their {\rd effect is small closer to the
Galactic centre}.

\item Model the dwarf as a series of Hernquist spheres that move on slightly
different orbits, so they are now spread roughly along the central orbit of
the family. A model along these lines would be significantly more realistic
than that presented here but almost as tractable. Since each component would
pass through pericentre at a different time, the overall effect would be the
extend the period over which the disc is disturbed, while holding the
aggregate effect roughly constant.

\item Include the effect on the disc of the wake that trails the dwarf as it
moves through the Galaxy's dark halo. The structure of such a wake has been
computed by \cite{Mulder1983}, \cite{Weinberg1993} and others.

\item Compute the impact that the dwarf has on the $(z,v_z)$ phase plane at a
given azimuth $\phi$ and radius $R$. This is a challenging task for two
reasons: (i) one has to work in the non-inertial frame of the barycentre of
the stars at $R$. This frame might reasonably be supposed to be frame defined
by the HT69 star at $(R,\phi)$. By definition the frame's acceleration
cancels the gravitational field at the particle's location that is jointly
generated by the dwarf and our Galaxy. What has to be computed is the
difference between the frame's acceleration and the gravitational field at
nearby points. This might be estimated by supposing that the middle of the
vertical profile of  a realistically thick
disk lies in the surface $z(R,\phi)$ computed here in the razor-thin
approximation.
 
\item {\rd Integrate the equations for $h(R)$ from $h=\dot h=0$ in the case that
the disc is embedded in a live dark halo that is out of equilibrium in the
sense that it comprises spheroidal shells, the symmetry axes of which point in
directions that shift as the shell's increases. \cite{HanConroy2023} presents an
investigation that is similar, but suffers from the crucial limitation of a
dead halo.}
\end{itemize}

Although an enormous amount of work is required before we can be sure that we
have solved the half-century old puzzle of warps, this re-visit to the ideas
of a classic paper strengthens the case that the Galactic warp is largely
generated by the dwarf's tidal field. When this idea has been put on a more
secure and quantitative basis, the warp phenomenon will surely provide
valuable probes of both dark haloes and stellar discs.

\section*{Acknowledgements}
It's a pleasure to thank Sacha Guerrini for clarifying the discussions that
clarified the work of Hunter \&
Toomre.
This research was supported in part by grant NSF PHY-2309135 to the Kavli
Institute for Theoretical Physics (KITP). 

\section*{Data Availability}

The computations were performed by C++ code linked to the AGAMA library
\citep{AGAMA}. The code is available upon request.

\bibliographystyle{mn2e} \bibliography{/u/tex/papers/mcmillan/torus/new_refs}

\appendix
{\rd 
\section{A subtle point about equation (5)}\label{sec:HTdiscuss}

Here we discuss a delicate point that was skated over in the derivation of
the equations of motion (\ref{eq:warpeqs}) in Section
\ref{sec:eqs}.  The problem is made apparent by computing the
potential $\Phi_0(R)$ of the  unperturbed axisymmetric disc. It is straightforward to
show that equation (\ref{eq:CohlTohlineeqsa}) yields
\[
\Phi_0(R)=-{2G\over\surd R}\int\d R'\,\surd R'\Sigma(R')Q_{-1/2}(\chi),
\]
so the vertical acceleration of the disc is
\begin{align}
a_{0z}&=-{\p\Phi_0\over\p z}
={2G\over\surd R}\int\d R'\,\surd R'\Sigma(R'){\d
Q_{-1/2}\over\d\chi}{z\over RR'},\cr
&={2Gz\over R^{3/2}}\int\d R'\,{\Sigma(R')\over\surd R'}{\d
Q_{-1/2}\over\d\chi}.
\end{align}
 As $z\to0$, the acceleration tends to a finite value despite the factor $z$ in front
of the integral: the integral diverges as $1/z$ because as $z$ diminishes,
$\chi\to1$ when $R'=R$ and the derivative of $Q$ diverges like
$1/(\chi-1)\simeq2(R/z)^2$. This finding calls into question the legitimacy
of pulling the derivative of $Q$ out of the $\phi'$ integral of equation
(\ref{eq:beforePull}). 

By modelling the disc as razor-thin we have created a difficult situation.
The vertical force changes sign at $z=0$ so it's not analytic there and
cannot be Taylor expended around $z=0$. Physically, the vertical force at
$(R,\phi,z)$ is strongly dominated by material in the disc that lies within a
circle of radius $z$ about $(R,\phi,0)$. Here the local
component of the force on a star is effectively the force between a
particle before and after displacement. We exclude this fictitious self-force
by excluding $R'=R$ from the integration over $R'$. That is, we break the
integral into parts over $(0,R-\epsilon_r)$ and $(R+\epsilon_r,\infty)$ and
let $\epsilon_r$ become small without vanishing. With $R'=R$ excluded, $\chi$
remains greater than unity even if $z-z'$ vanishes. Moreover, when
$|z-z'|\ll\epsilon_r$, $Q$ and its derivatives become independent of $z-z'$
and therefore independent of $\phi'$, so it's legitimate to pull $\d
Q/\d\chi$ out of the integral in equation (\ref{eq:beforePull}). 

Subtly different equations of motion are obtained in HT69 by writing an
expression for $a_z$ directly (eq 2 in HT69) rather than obtaining $a_z$ from
the potential. Appendix \ref{sec:appHTeqs} reproduces the derivation in HT69 using
cylindrical rather than oblate-spheroidal coordinates. The integral of
equation (\ref{eq:azfrommu}) is stated to be the force from a sheet of dipoles
with surface density proportional to $h-h'$. Actually the expression is only
one of the two terms required for the force from a dipole. Appendix
\ref{sec:appHTeqs} makes a case that the other term can be neglected, but
the case is delicate and HT69 does not attempt to make it. 

In HT69 the contributions to $a_z$ from $h(R')$ and $h(R)$ are separated. That
proportional to $h(R)$ is found to be associated with two discs with equal but opposite
surface densities that are infinitesimally displaced in $z$, which is shown
to be
 \[
a_{2z}=h(R,\phi){1\over R}{\p v_c^2\over\p
R}.
\]
The other contribution is found by solving for the
potential of a more complex sheet of dipoles using oblate confocal spheroidal
coordinates. An analogous solution using equations (\ref{eq:CohlTohlineeqsa})
yields the Fourier coefficients
\begin{align}\label{eq:withDa}
a_{1m}(R)&=-\lim_{z\to0}{\p\Phi_m\over\p z}\cr
&=-{2G\over R^{3/2}}\int\d R'\,{D_m(R,R')\over\surd R'}\Sigma(R')h_m(R'),
\end{align}
where
\begin{align}\label{eq:withDb}
D_m(R,R')&\equiv\lim_{z\to0}\bigg({\d Q_{|m|-1/2}\over\d\chi}
+{z^2\over RR'}{\d^2 Q_{|m|-1/2}\over\d\chi^2}
\bigg)\bigg|_{z'=0}.
\end{align}
 Thus we encounter the same problem as before when $R'$ passes through $R$,
causing $Q$ and its derivatives to diverge as $z\to0$. By
dividing the integral into parts of $(0,R-\epsilon_r)$ and
$(R+\epsilon_r,\infty)$ we can argue that the second term in equation
(\ref{eq:withDb}) can be neglected because its contributions are more tightly confined
around $R=R'$ than those of the first term. Then equation (\ref{eq:withDa}) yields
essentially the first term in the integral of equation (\ref{eq:azeqa}).
HT69 does not regularise the integral over $D_m$ as we have
done in equation (\ref{eq:azeqa}) by subtracting a term proportional to $h(R,\phi)$,
so it  arrives at equations that make sense only for a finite number of
discrete rings. 

Thus if extended to a continuum, the equations of motion in HT69 contain one divergent
integral and a finite term. By contrast with the counter terms in equation
(\ref{eq:azeqb}) deleted we have two divergent integrals, and by inclusion of the
counter terms both integrals are regularised.
With the counter term deleted, equation (\ref{eq:azeqb}) yields an integral for
$RN(R)$ that coincides with 
\[\label{eq:Dagain}
{2G\over \surd R}\int\d R'\,{D_0(R,R')\over\surd
R'}\Sigma(R')
\]
 when the double derivative is dropped from the definition (\ref{eq:withDb}) of
$D_0$.  Numerical evaluations show that the integral (\ref{eq:Dagain}) yields $\d
v_c^2/\d R$ only when the second derivative is included in $D_0$ -- the
divergent contributions of the two terms in $D_0$ cancel to produce the finite value
that HT69 gives for the coefficient of $h(R)$. In our
treatment the integral over $\d Q_{-1/2}/\d\chi$ is regularised by an integral
over $\d Q_{|m|-1/2}/\d\chi$ rather than an integral over
$\d^2Q_{-1/2}/\d\chi^2$ and the coefficient of $h(R)$ in the equations of
motion isn't $(\d v_c^2/\d R)/R$.

Decreasing the parameter $\epsilon_z$ improves energy conservation at the
cost of increased computational cost. In the computations below we have set
$\epsilon_r=10^{-3}R$ and $\epsilon_z=10^{-4}R$, where $R$ is the radius at
which the potential is required. With this choice the \rms\ variation in
energy during the evolution shown in Fig.~\ref{fig:freeWarp} is $2.5$ per
cent.  

An appendix of \cite{SparkeCasertano1988} gives an insightful derivation of
alternative expressions for vertical accelerations $a_{1z}$ and $a_{2z}$. In
these expressions $a_{1z}$ and $a_{2z}$ are obtained by integrating products of
$h_m$ and the kernel
\[
B_m(R,R')\equiv\lim_{z\to0}\int_0^\infty\d k\,k^2J_m(kR)J_m(kR')\,\e^{-kz},
\]
where $J_m$ is the usual cylindrical Bessel function. As $z\to0$ the integral over $k$
becomes costly to evaluate on account of the oscillating nature of $J_m$.
The Greens' function (\ref{eq:CohlTohlineeqsa}) provides an explicit formula for  the analogous
kernel, which moreover is not an oscillating function. Hence  costly
integrations are eliminated. 
}
\section{Vertical forces according to HT69}\label{sec:appHTeqs}

We adapt the approach to computing vertical forces  taken in HT69 to the use of
the Greens' function (\ref{eq:CohlTohlineeqsa}).
The gravitational acceleration at $\vx$ due to the mater at
$\vx'$ is 
\[
\d\va(\vx)={G\Sigma(|\vx'|)\d^2\vx'\over|\vx'-\vx|^3}(\vx'-\vx).
\]
We are only interested in the vertical component of this acceleration. Since
$\vx'$
lies above the point $\vx$ by $h(\vx')-h(\vx)$,  the vertical component is
\[\label{eq:azfrommu}
\d a_z(\vx,\vx')=
{G\Sigma(|\vx'|)\d^2\vx'\over|\vx'-\vx|^3}[h(\vx')-h(\vx)].
\]

The gravitational acceleration $\va(\vp,\vx,\vx')$ at $\vx$ from a dipole $\vp=m\vdelta$ at $\vx'$ is
obtained by differencing the accelerations towards masses $\pm m$ located at
$\vx'\pm\frac12\vdelta$. 
\begin{align}\label{eq:adipole}
\va(\vp,\vx,\vx')&=G\vp\cdot{\p\over\p\vx'}{\vx'-\vx\over|\vx'-\vx|^3}\cr
&=G\bigg({\vp\over|\vx'-\vx|^3}
-3{\vp\cdot(\vx'-\vx)\over|\vx'-\vx|^5}(\vx'-\vx)\bigg).
\end{align}
Now consider the difference in the accelerations produced by two
distributions of dipoles over a disc. Let the dipoles  be 
\begin{align}\label{eq:twoDipoles}
\d\vp_1(\vx')&=\Sigma(|\vx'|)h(\vx')\d^2\vx'\,\ve_z\cr
\d\vp_2(\vx')&=\Sigma(|\vx'|)h(\vx)\d^2\vx'\,\ve_z.
\end{align}
Then the difference in the accelerations they produce at $\vx=(R,h(\vx),\phi)$ is
\begin{align}\label{eq:diffais}
\va_1-\va_2&=G\int\d^2\vx'\,\Sigma(|\vx'|)\bigg({h(\vx')-h(\vx)\over|\vx'-\vx|^3}\ve_z\cr
&\qquad-3\cos\theta{|h(\vx')-h(\vx)|\over|\vx'-\vx|^4}(\vx'-\vx)\bigg),
\end{align}
 where $\theta$ is the angle between $\vx'-\vx$ and $\ve_z$.  As $h\to0$, the
area of the disc within which this angle is not close to $\pi/2$ shrinks,
which suggests that in this limit the second term in the integrand can be
neglected. A contrary consideration is that this term grows faster than the
other term by one
power of $|\vx-\vx'|^{-1}$ precisely where $\cos\theta$ is non-negligible.
HT69 implicitly assumes that the integral becomes dominated by the first term,
which yields the required vertical acceleration
\eqref{eq:azfrommu} and thus argue that the vertical
acceleration produced by a warped disc can be obtained as the difference
between the vertical accelerations produced by the dipole distributions
\eqref{eq:twoDipoles}.

The acceleration $a_{2z}$ can be evaluated from the unperturbed potential
$\Phi_0$ because it is just the acceleration due to two discs that differ
only in the sign of $\Sigma$ and are vertically separated by the constant
$h(\vx)$. Hence
 \[
a_{2z}(\vx)=h(\vx){\p^2\Phi_0\over\p z^2}\bigg|_\vx,
\]
 where $z$ lies just above the disc. Here the density vanishes and Poisson's
equation yields
\[
{\p^2\Phi_0\over\p z^2}=-{1\over R}{\p v_c^2\over\p R},
\]
so\footnote{Notice that $a_{2z}/h$ is negative near the centre where $v_c$ is
rising. The integral over the first term in our expression \eqref{eq:adipole}
for a dipole's acceleration inevitably makes a positive contribution to
$a_z/h$, so the negativity of $a_{2z}/h$ is caused by the second term in
\eqref{eq:adipole}, which does not appear in HT69.}
\[\label{eq:aztwo}
a_{2z}(\vx)=-{h(\vx)\over R}{\p v_c^2\over\p R}.
\]
This acceleration depends on $z$ only through $h(\vx)$, which is the
amplitude of the dipole sheet; the acceleration above a given dipole sheet is
independent of $z$.

We now compute the acceleration $a_{1z}$ produced by the dipole distribution
$\d\vp_1$ in equation \eqref{eq:twoDipoles}. We expand the displacement field
$h(R,\phi)$ in the Fourier series (\ref{eq:hFourier}) and use equation
(\ref{eq:CohlTohlineeqsa}) to compute the contribution to $a_{1z}$ from the
dipole density $p(R,\phi)=\Sigma(R)h_m(R)\e^{\i m\phi}$. Inserting the
density
$\rho(R,z,\phi)=\sigma(R,\phi)\delta(z-\zeta)$ of a disc that lies in the
plane $z=\zeta$, we obtain the Fourier coefficients
\[\label{eq:discPhiQb}
\Phi_m(R,z)\equiv-{2G\over\sqrt{R}}
\int\d R'\,\sqrt{R'}\sigma_m(R')Q_{|m|-1/2}(\chi)
\]
with $\chi$ is evaluated with $z'=\zeta$ and
\[
\sigma_m(R')\equiv\int_0^{2\pi}{\d\phi'\over2\pi}\,\sigma(R',\phi')\e^{\i m\phi'}.
\]
The potential jointly generated by  two razor-thin
discs that carry opposite signs of mass and are separated by a small distance
$2\zeta$ has Fourier coefficients
 \begin{align}\label{eq:Phidipole}
\Phi_m(R,z)&=-{2G\over\surd R}\int\d
R'\,\surd
R'\sigma_m(R')\big[Q_{|m|-1/2}(\chi(z'=\zeta))\cr
&\hskip2.5cm -Q_{|m|-1/2}(\chi(z'=-\zeta))\big]\cr
&={2G\over\surd R}\int\d
R'\,\surd
R'p_m(R'){\d Q_{|m|-1/2}\over\d\chi}\bigg|_{z'=0}{z\over RR'},
\end{align}
where $p_m(R)=2\zeta\sigma_m(R)$ is the Fourier component of the dipole
density represented by the two sheets.
$Q_{|m|-1/2}(\chi)$ diverges like $\ln(\chi-1)$ as $\chi\to1$. Consequently,
as $\chi\to1$ the derivative in equation \eqref{eq:Phidipole} diverges like
$1/(\chi-1)$ so as $R\to R'$ and $z\to0$ the product of $z$ and the
derivative of $Q$ diverges. Nonetheless, the integral over $R'$ at nonzero
$z$ tends to a finite limit as $z\to0$.  The limit's value is the potential
just above the dipole sheet. The potential below the sheet is minus that
value.

To obtain the vertical accelerations that we need, we set
$p_m(R')=\Sigma(R')h_m(R')$ in equation (\ref{eq:Phidipole})
and compute
\begin{align}
\label{eq:azoneb}
a_m(R')&=-\lim_{z\to0}{\p\Phi_m\over\p z}\cr
&=-{2G\over R^{3/2}}\int\d R'\,{D(R')\over\surd R'}\Sigma(R')h_m(R'),
\end{align}
where
\begin{align}
D(R,R')&\equiv\lim_{z\to0}{\p\over\p z}\bigg(z{\d Q_{|m|-1/2}\over\d\chi}
\bigg)\bigg|_{z'=0}.
\end{align}

\section{Energy integrals}\label{sec:appE}

With $h(R,\phi,t)$ the distance of the disc above the plane $z=0$, the
equation of motion of a star is
\begin{align}
{\p^2 h\over\p t^2}+2\Omega(R){\p^2 h\over\p t\p\phi}+\Omega^2{\p^2 h
\over\p\phi^2}=G\int\d^2\vR'\,\Sigma(R'){h(\vR')-h(\vR)\over|\vR-\vR'|^3}.
\end{align}
We multiply by $\Sigma(R)(\p h/\p t)\d^2\vR$ and integrate
\begin{align} \label{eq:aftermult}
\int\d^2\vR\,\Sigma&\bigg(
{\p^2 h\over\p t^2}{\p h\over\p t}+2\Omega{\p^2 h\over\p t\p\phi}{\p
h\over\p t}+\Omega^2{\p^2h\over\p\phi^2}{\p h\over\p t}\bigg)
=G\int\d^2\vR\cr
&\times\int\d^2\vR'\,\Sigma(R)\Sigma(R'){h(\vR')-h(\vR)\over|\vR-\vR'|^3}{\p
h(\vR)\over\p t}.
\end{align}
The middle term on the left can be rearranged to
\[
\Omega(R){\p\over\p\phi}\bigg({\p h\over\p t}\bigg)^2,
\]
which vanishes when integrated over $\phi$. After using
$(\p^2h/\p\phi^2)=-m^2h$, the derivative w.r.t.\ $t$ can be taken
out of the remaining integral. The right side of equation
\eqref{eq:aftermult} only changes its sign when the primed and unprimed variables are
swapped. So we can subtract its swapped version and halve the result, to
obtain
\begin{align} \label{eq:genHTint}
0=&{\d\over\d t}\bigg\{\fracj12\int\!\d^2\vR\,\Sigma\bigg(
\Big({\p h\over\p t}\Big)^2-m^2\Omega^2h^2\bigg)
\cr
&+{G\over4}\int\!\d^2\vR\int\!\d^2\vR'\,\Sigma(R)\Sigma(R')
{[h(\vR')-h(\vR)]^2\over|\vR-\vR'|^3}\bigg\},
\end{align}
 which gives us an energy integral for each $m$.

When $m=0$, this integral provides a proof of stability because then the
potential energy comprises the manifestly non-negative double integral. When
$m\ne0$ the single integral contains a negative contribution to the potential
energy and stability is not evident. For the case $m=1$ Hunter \& Toomre were able to demonsrate
stability by combining the potential energy terms into a single non-negative
double integral as follows.  

Poisson's integral for the potential of the unperturbed disc is 
\[
\Phi(R)=-\int\d^2\vR'\,{G\Sigma(R')\over\sqrt{R^2+R^{\prime2}-2RR'\cos(\phi'-\phi)}}.
\]
Differentiating we get
\[
\Omega^2(R)={1\over R}{\p\Phi\over\p R}
=G\int\d^2\vR'\,\Sigma(R'){1-(R'/R)\cos(\phi'-\phi)\over|\vR-\vR'|^3}.
\]
Inserting this into equation \eqref{eq:aftermult} with $m=1$, we find
\begin{align}\label{eq:nearlyE}
{\d\over\d t}\fracj12\int\d^2\vR\,&\Sigma\bigg(
{\p h\over\p t}\bigg)^2=G\int\d^2\vR\int\d^2\vR'\,\Sigma(R)\Sigma(R')\cr
&\times{h(\vR')-h(\vR)(R'/R)\cos(\phi'-\phi)\over|\vR-\vR'|^3}{\p
h(\vR)\over\p t}.
\end{align}
In the double integral over angles on the right we change the inner
integration variable from $\phi'$ to $\Delta\equiv\phi'-\phi$, which already
occurs in both the numerator and denominator. Now
\begin{align}
&h(\vR')=h(R',\phi')=h(R',\phi+\Delta)\cr
&=h_{\rm c}\cos(\phi+\Delta)+h_{\rm s}\sin(\phi+\Delta)\cr
&=h_{\rm c}\big[\cos\phi\cos\Delta-\sin\phi\sin\Delta\big]
+h_{\rm s}\big[\sin\phi\cos\Delta+\cos\phi\sin\Delta\big]\cr
&=\big[h_{\rm c}\cos\phi+h_{\rm s}\sin\phi\big]\cos\Delta
+\big[-h_{\rm c}\sin\phi+h_{\rm s}\cos\phi\big]\sin\Delta\cr
&=h(R',\phi)\cos\Delta+h(R',\phi+\pi/2)\sin\Delta.
\end{align}
When we replace $h(\vR')$ in the numerator of equation \eqref{eq:nearlyE} with this
expression, the term proportional to $\sin\Delta$ vanishes on integration
over $\Delta$ because it's an odd function of $\Delta$ while the
denominator  is an even  function so the integrand is odd. After the
replacement we therefore have
\begin{align} \label{eq:dEdtHT}
{\d\over\d t}&\fracj12\int\d^2\vR\,\Sigma\bigg(
{\p h\over\p t}\bigg)^2=G\int\d R\,R\Sigma(R)\int\d R'\,R'\Sigma(R')\cr
&\times\int\d\phi\,
\big[h(R',\phi)-h(R,\phi)(R'/R)]{\p h(\vR)\over\p t}
\int_0^{2\pi}\d\Delta\,{\cos\Delta\over|\vR-\vR'|^3}\cr
&=-{G\over4}{\d\over\d t}\int\d R\,\Sigma(R)\int\d
R'\Sigma(R')\cr
&\hskip1cm\times\int\d\phi\big[Rh(R',\phi)-R'h(R,\phi)\big]^2
\int_0^{2\pi}\d\Delta{\cos\Delta\over|\vR-\vR'|^3}.
\end{align}
The argument of the time derivative on the right is now manifestly
non-negative, which establishes stability except when the argument is
identically zero, which it is when $Rh(R',\phi)=R'h(R,\phi)$. This condition
is satisfied when the perturbation amounts to a tilt of the whole disc about
an axis that lies in the plane -- it's physically obvious that such tilts
should involve no potential energy and thus be neutrally stable.

The integral over $\Delta$ in equation (\ref{eq:dEdtHT}) simplifies:
\begin{align}
\int\d\Delta{\cos\Delta\over|\vR-\vR'|^3}&=
\int\d\Delta{\cos\Delta\over[R^2+R^{\prime2}-2RR'\cos\Delta]^{3/2}}\cr
&={g(\chi)\over
[R^2+R^{\prime2}]^{3/2}},
\end{align}
where $\chi$ is defined by equation \eqref{eq:CohlTohlineeqsb} and
\[
g(\chi)\equiv2\int_0^{\pi}\d\Delta\,{\cos\Delta\over(1-\chi^{-1}\cos\Delta)^{3/2}}.
\]
 The integral over $\phi$ in equation \eqref{eq:dEdtHT} can be executed analytically. We
use that $h(\phi)=2(h_{\rm r}\cos\phi-h_\i\sin\phi)$, where $h_{\rm r}$ and $h_\i$
are the real and imaginary parts of $h$. Then
\begin{align}
\fracj14\int&\d\phi\big[Rh(R',\phi)-R'h(R,\phi)\big]^2\cr
&=\int\d\phi\Big\{R\big[h_{\rm r}(R')\cos\phi-h_\i(R')\sin\phi\big]\cr
&\hskip2.5cm-R'\big[h_{\rm r}(R)\cos\phi-h_\i(R)\sin\phi\big]\Big\}^2\cr
&\hskip-.6cm=\pi[Rh_{\rm r}(R')-R'h_{\rm r}(R)]^2+\pi[Rh_{\rm i}(R')-R'h_{\rm i}(R)]^2.
\end{align}
With these results, equation \eqref{eq:dEdtHT} becomes the statement that
for $m=1$ the conserved energy is
 \begin{align}\label{eq:finalHTintegral}
&E=2\pi\int\d R\,R\Sigma(R)\bigg[\Big({\p h_{\rm r}\over\p t}\Big)^2\cr
&\hskip.8cm+\Big({\p h_\i\over\p t}\Big)^2\bigg]
+\pi G\int\d R\,\Sigma(R)\int\d R'\,{\Sigma(R')g(\chi)\over [R^2+R^{\prime2}]^{3/2}}\cr
&\hskip-.2cm\times\Big\{
[Rh_{\rm r}(R')-R'h_{\rm r}(R)]^2+[Rh_{\rm i}(R')-R'h_{\rm
i}(R)]^2\Big\}.
\end{align}
The double integral is symmetric in $R,R'$ so is equal to twice the  integral over
$R'<R$. In terms of $s\equiv\ln(R'/R)$ we have
 \begin{align}%\label{eq:finalHTintegral}
PE&=2\pi G\int_0^\infty\d R\,\Sigma(R)\int_{-\infty}^0\d s\,\Sigma(R')
{\e^sg(\chi)\over [1+\e^{2s}]^{3/2}}\cr
&\times\Big\{
[h_{\rm r}(R')-\e^sh_{\rm r}(R)]^2+[h_{\rm i}(R')-\e^sh_{\rm
i}(R)]^2\Big\}.
\end{align}
 As $R\to R'$, $g(\chi)$
diverges but the integrand remains finite because the
curly bracket then tends to zero.

\end{document}